\documentclass[twocolumn,superscriptaddress,aps,pre]{revtex4-1}
\usepackage{amssymb,amsmath}
\usepackage{graphicx}
\usepackage[utf8]{inputenc}
\usepackage[colorlinks,linkcolor=blue,citecolor=blue,urlcolor=blue]{hyperref}

\begin{document}

\title{Probability density of the fractional Langevin equation with reflecting walls}
\author{Thomas Vojta}
\affiliation{Department of Physics, Missouri University of Science and Technology, Rolla, MO 65409, USA}

\author{Sarah Skinner}
\affiliation{Department of Physics, Missouri University of Science and Technology,
Rolla, MO 65409, USA}

\author{Ralf Metzler}
\affiliation{Institute of Physics and Astronomy, University of Potsdam, D-14476 Potsdam-Golm, Germany}

\begin{abstract}
We investigate anomalous diffusion processes governed by the fractional Langevin equation and
confined to a finite or semi-infinite interval by reflecting potential barriers.
As the random and damping forces in the fractional Langevin equation fulfill the appropriate fluctuation-dissipation relation,
the probability density on a finite interval converges for long times towards the expected
uniform distribution prescribed by thermal equilibrium.
In contrast, on a semi-infinite interval with a reflecting wall at the origin, the
probability density shows pronounced deviations from the Gaussian behavior observed for normal
diffusion. If the correlations of the random force are persistent (positive), particles
accumulate at the reflecting wall while anti-persistent (negative) correlations lead to
a depletion of particles near the wall.
We compare and contrast these results with the strong accumulation and depletion effects recently
observed for non-thermal fractional Brownian motion with reflecting walls, and we discuss broader implications.
\end{abstract}

\date{\today}

\maketitle

\section{Introduction}
\label{sec:Intro}

Normal diffusion is an almost omnipresent transport process that can be found in biological,
chemical, as well as physical systems, and beyond. It is characterized by a linear relation
between the mean-square displacement $\langle x^2 \rangle$ of the diffusing object and
the elapsed time $t$. According to Einstein \cite{Einstein_book56}, diffusive behavior
results from stochastic motion that is local in both time and space. This means individual
random displacements (step lengths) have a finite correlation time and a finite variance.

In recent years, considerable attention has been attracted by anomalous diffusion,
i.e., random motion that does not fulfill the dependence $\langle x^2 \rangle \sim t$.
Both subdiffusion (for which $\langle x^2 \rangle$ grows slower than $t$) and
superdiffusion (where $\langle x^2 \rangle$ grows faster than $t$) have been
observed in numerous experiments and described mathematically
(for reviews see, e.g., Refs.\
\cite{MetzlerKlafter00,HoeflingFranosch13,BressloffNewby13,MJCB14,MerozSokolov15,MetzlerJeonCherstvy16,Norregaardetal17}
and references therein).

Anomalous diffusion can arise via several different mechanisms if the condition of
locality in space and time is violated. For example, if the step lengths are broadly distributed
such that their variance does not exist, the motion can become superdiffusive. Conversely,
a broad distribution of waiting times with a diverging mean may lead to subdiffusive motion.
Another important mechanism that leads to anomalous diffusion is the presence of
long-range power-law correlations in time between individual random displacements (steps).
Such correlations can produce subdiffusive or superdiffusive behavior even if the
step lengths and waiting times are narrowly distributed. Fractional Brownian motion
(FBM) and the fractional Langevin equation (FLE)  are two prototypical mathematical
models for this situation.

FBM was introduced by Kolmogorov \cite{Kolmogorov40} and further investigated by Mandelbrot and van Ness \cite{MandelbrotVanNess68}.
It has been studied in the mathematical literature quite extensively (see, e.g., Refs.\ \cite{Kahane85,Yaglom87,Beran94,BHOZ08}),
and it has been applied to a wide variety of problems including polymer dynamics \cite{ChakravartiSebastian97,Panja10}, diffusion inside living cells \cite{SzymanskiWeiss09,Jeonetal11}, traffic in electronic networks \cite{MRRS02}, as well as stock markets \cite{ComteRenault98,RostekSchoebel13}.
FBM is a self-similar, non-Markovian Gaussian process with stationary increments $\xi$ that feature long-range power-law correlations
in time.  If the increments are positively correlated (persistent), the resulting motion is superdiffusive whereas
anticorrelated (antipersistent) increments produce subdiffusive motion.

Recently, large-scale computer simulations of FBM confined by reflecting walls have demonstrated that
the interplay between the long-time correlations and the confinement strongly affects the probability density
function $P(x,t)$ of the diffusing particle.  Specifically, if the motion is restricted to a semi-infinite
interval by a reflecting wall at the origin, the probability density becomes highly non-Gaussian and develops
a power-law singularity, $P \sim x^\kappa$ at the wall \cite{WadaVojta18,WadaWarhoverVojta19}. For persistent noise,
particles accumulate at the wall, $\kappa<0$. In contrast, particles are depleted close to the wall, $\kappa > 0$ for
anti-persistent noise. Analogous simulations of FBM
on a finite interval, with reflecting walls at both ends, have established that the stationary probability density
reached for long times strongly deviates from the uniform distribution found for normal diffusion
\cite{Guggenbergeretal19}. Particles accumulate at the walls and are depleted in the middle of the interval
for persistent noise whereas the opposite is true for anti-persistent noise.

FBM can be understood as random motion governed by \emph{external} noise
\cite{Klimontovich_book95}. It does not obey the fluctuation-dissipation
theorem and generally does not reach a thermal equilibrium state. It is thus interesting to ask whether
the unusual features of the probability density close to reflecting walls survive or disappear if the
long-range correlated noise fulfills the fluctuation-dissipation theorem.

To answer this question, we study in this paper the fractional Langevin equation (FLE) in the presence of reflecting
potentials that restrict the motion to finite or semi-infinite intervals. Our results can be summarized
as follows. As the random and damping forces fulfill the appropriate fluctuation-dissipation relation,
the FLE motion confined to a finite interval is expected to approach the thermal equilibrium state for
long times. We show here that this is indeed the case. Consequently, the probability density of the particle position
approaches a uniform distribution.
In contrast, the probability density on  a semi-infinite interval shows pronounced deviations from the
Gaussian behavior observed for normal diffusion. If the correlations of the random force are persistent, particles
accumulate at the reflecting wall while anti-persistent correlations lead to a depletion of particles near the wall,
compared to the naively expected Gaussian distribution.

In addition, we analyze a generalized Langevin equation featuring long-range correlated fractional
noise but a conventional, instantaneous damping term. For this equation, which violates the fluctuation-dissipation
theorem, we find that the probability density on a finite interval is non-uniform and resembles the result
for FBM.

Finally, we compare reflecting potentials (as used in the bulk of our work) with hard reflecting boundary
conditions for which the velocity simply changes sign when the particle hits the wall, $v \to -v$.
Such reflecting boundary conditions were recently employed by Holmes \cite{Holmes19} who reported a
non-uniform stationary probability density for the FLE on a finite interval. Our simulations demonstrate that
the FLE with hard reflecting boundary conditions is very sensitive to the time step in the numerical integration.
A uniform distribution, as required by thermal equilibrium, is only recovered for extremely small time steps.

The remainder of our paper is organized as follows. We introduce the FLE in Sec.\ \ref{sec:FLE} where we also describe our simulation method. Section \ref{sec:finite} is devoted to the results for the FLE on a finite interval, whereas
Sec.\ \ref{sec:semi-infinite} presents the behavior for a semi-infinite interval.
Anti-persistent noise and damping forces that violate the fluctuation dissipation theorem  are
discussed in Sec.\ \ref{sec:extensions}.  We conclude in Sec.\ \ref{sec:conclusions}.
The appendix is devoted to a discussion of the boundary conditions as well as an algorithm
to speed up the solution of the FLE.

\section{Fractional Langevin equation}
\label{sec:FLE}
\subsection{Definition}

Our starting point is the well-known Langevin equation \cite{Langevin08},
\begin{equation}
m \frac {d^2}{dt^2} x(t) = - \bar\gamma \frac d {dt} x(t) + \xi_w(t)~.
\label{eq:LE}
\end{equation}
It describes the time evolution of the position $x$ of a particle of mass $m$ moving under
the influence of an uncorrelated random force (Gaussian white noise) $\xi_w(t)$  and a linear
damping force with damping coefficient $\bar\gamma$.

If the random force is correlated (nonwhite noise) and the damping force is nonlocal in time,
the motion follows the generalized Langevin equation \cite{Zwanzig_book01,Haenggi78,Goychuk12}
\begin{equation}
m \frac {d^2}{dt^2} x(t) = - \bar\gamma \int_{0}^t dt' \mathcal K(t-t') \frac d {dt'} x(t') + \xi(t)~.
\label{eq:GLE}
\end{equation}
In thermal equilibrium at temperature $T$, the fluctuation-dissipation theorem \cite{Kubo66}
requires that the noise correlation function and the damping (memory) kernel $\mathcal K$ are related via
\begin{equation}
\langle \xi(t) \xi(t') \rangle = k_B T \bar\gamma \mathcal K(t-t')~.
\label{eq:FDT}
\end{equation}
We are interested in the case of $\xi(t)$ being a fractional Gaussian noise \cite{Qian03},
i.e., a Gaussian process of zero mean, $\langle \xi(t) \rangle=0$ and a power-law correlation function
\begin{equation}
\langle \xi(t) \xi(t') \rangle \sim \alpha(\alpha-1)K_\alpha |t-t'|^{\alpha-2}
\label{eq:FGN}
\end{equation}
for $t \ne t'$.
In this case, eq.\ (\ref{eq:GLE}) is called the FLE \cite{Lutz01}. Here,
$K_\alpha$ denotes the noise amplitude. The correlation exponent $\alpha$ is restricted
to the range $1 <\alpha <2$ because the damping integral in (\ref{eq:GLE})
diverges at $t=t'$ for $\alpha < 1$, and $\alpha > 2$ is unphysical as it corresponds to correlations that
increase with time. (The Hurst exponent $H$, often used in the mathematical literature,
is given by $H=\alpha/2$.) In the range $1 <\alpha <2$, the fractional Gaussian noise is persistent (positively
correlated). The limiting case, $\alpha=1$, corresponds to normal diffusion.

The behavior of the FLE in the absence of confining potentials is well understood (see, e.g., Ref.\
\cite{MJCB14} and references therein).  Consider, e.g., a particle starting from rest at the origin at
time $t=0$. The probability densities of both its velocity and its position are Gaussian.
For long times, the mean-square velocity approaches the value $k_B T/m$ prescribed by the
classical equipartition theorem whereas the mean-square displacement crosses over
from ballistic behavior, $\langle x^2(t) \rangle \sim t^{2}$, at short times to anomalous diffusion,
\begin{equation}
\langle x^2(t) \rangle \sim t^{2-\alpha}~,
\label{eq:MSD_free}
\end{equation}
at long times. This means that the FLE with persistent noise, $1 <\alpha <2$, leads to subdiffusion while
FBM with the same noise produces superdiffusion, $\langle x^2(t) \rangle \sim t^\alpha$.
This is caused by the fact that the fluctuation-dissipation relation (\ref{eq:FDT}) couples
persistent noise with long-range memory in the damping term.

To confine the motion to either a finite or a semi-infinite interval, we introduce
reflecting walls via repulsive potentials. Specifically, to model a reflecting wall
at position $x_0$, we introduce an external potential
\begin{equation}
V(x)= V_0 \exp[\mp \lambda (x-x_0)]~.
\label{eq:wall_potential}
\end{equation}
The $\pm$ sign in the exponent distinguishes walls reflecting
from the right or from the left. This potential creates an additional external force
\begin{equation}
F(x)= -dV(x)/dx = \pm F_0 \exp[\mp \lambda (x-x_0)]
\label{eq:wall_force}
\end{equation}
with $F_0=\lambda V_0$ on the r.h.s.\ of the FLE (\ref{eq:GLE}).  We are interested
in the limit of large $\lambda$ for which the decay length $\lambda^{-1}$ of the wall
force is small compared to the considered interval lengths $L$. In this limit, particles
in the interior of the interval will not experience an external force.
The slight ``softening'' of the reflecting wall due to a finite $\lambda$ is needed
for the consistent numerical integration of the FLE as described in Sec.\
\ref{subsec:simulation}.
For more details and a comparison with alternative implementations of reflecting walls in the FLE,
see Appendix \ref{sec:appendix_A}.

\subsection{Simulation method}
\label{subsec:simulation}

To simulate the FLE (\ref{eq:GLE}), we set the mass $m$, the damping coefficient
$\bar\gamma$, and the Boltzmann constant $k_B$ to unity. We then discretize time $t_n =n \Delta t$ with $n=0, 1, \ldots, N_t$.
After replacing the time derivatives with finite-difference expressions, the FLE turns
into the recursion relations
\begin{eqnarray}
v_{n+1} &=& v_n + \Delta t \left[\xi_n + F(x_n) - \sum_{m=0}^n {\mathcal K}_{n-m} v_m \right]~,
\label{eq:FLE_discrete_v}
\\
x_{n+1} &=& x_n + \Delta t\, v_n ~.
\label{eq:FLE_discrete_x}
\end{eqnarray}
The $\xi_n$ are a discrete fractional Gaussian noise \cite{Qian03}, i.e.,
identical Gaussian random numbers of zero mean and correlation function
\begin{equation}
\langle \xi_m \xi_{m+n} \rangle = K_\alpha (\Delta t)^{\alpha-2} \left( |n-1|^\alpha - 2 |n|^\alpha + |n+1|^\alpha \right)~.
\label{eq:FGN_discrete}
\end{equation}
These correlated random numbers are calculated using the Fourier-filtering method \cite{MHSS96}.
It starts from a sequence of independent Gaussian random numbers $\chi_n$. The Fourier transform $\tilde \chi_\omega$ of these numbers
is then converted via ${\tilde{\xi}_\omega} = [\tilde C(\omega)]^{1/2} \tilde{\chi}_\omega$,
where $\tilde C(\omega)$ is the Fourier transform of the correlation function (\ref{eq:FGN_discrete}).
The inverse Fourier transformation of the ${\tilde{\xi}_\omega}$ gives the desired fractional Gaussian noise.

The noise and the damping kernel fulfill the discrete version of the fluctuation-dissipation relation
\begin{equation}
\langle \xi_m \xi_{m+n} \rangle =  T {\mathcal K}_n~.
\label{eq:FDT_discrete}
\end{equation}

A naive implementation of the recursion (\ref{eq:FLE_discrete_v}) is numerically
expensive because the evaluation of a single damping integral scales linearly with the
number of time steps. Thus the total effort grows quadratically with the number
of time steps. We have devised an improved algorithm that speeds up the evaluation
of the damping integrals by several several orders of magnitude. This algorithm is
discussed in Appendix \ref{sec:appendix_B}.

\begin{figure}
\includegraphics[width=\columnwidth]{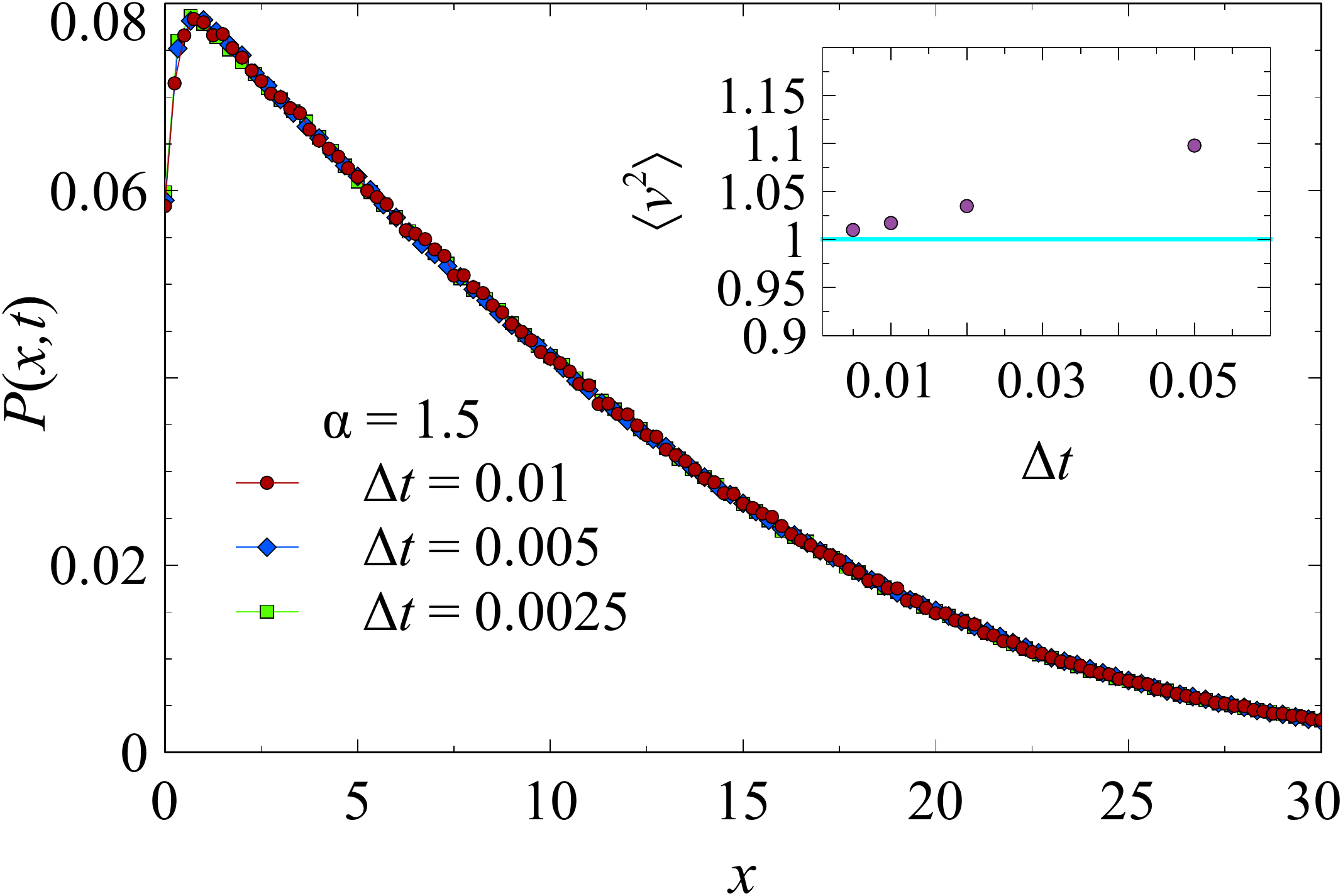}
\caption{Determination of a suitable value of the time step $\Delta t$.
Main panel: probability density $P(x,t)$ of the particle position at time $t=20972$ for correlation exponent $\alpha=1.5$, temperature
$T=1$, and several values of the time step $\Delta t$ (the particles start at the origin at $t=0$).
Inset: Mean-square velocity $\langle v^2 \rangle$ at long times as a function of time step $\Delta t$
(the data are averages of $\langle v^2 \rangle$ over the time intervals $10^4 - 10^5$ or
$10^4 - 10^6$ depending on $\Delta t$). }
\label{fig:v2vsdt}
\end{figure}

Choosing a suitable value for the time step $\Delta t$ is crucial for the
performance of the simulation. On the one hand, $\Delta t$ needs to be small enough
to limit the error due to the time discretization
\footnote{The choice of $\Delta t$ is affected by the magnitude and steepness of
the wall force (\ref{eq:wall_force}) with larger values of $F_0$ and $\lambda$
requiring smaller $\Delta t$.}.
 On the other hand, a small
$\Delta t$ increases the numerical effort to reach long simulation times.
To optimize the time step, we study the dependence on $\Delta t$ of the
mean-square velocity $\langle v^2 \rangle$ at long times of the FLE restricted to the semi-infinite
interval $(0,\infty)$.
The inset of Fig.\ \ref{fig:v2vsdt} shows $\langle v^2 \rangle$ vs.\ $\Delta t$ for $\alpha=1.5$ at temperature
$T=1$.
The data show that the deviation of $\langle v^2 \rangle$ from the value of 1 (required by the
fluctuation-dissipation theorem) decreases with decreasing $\Delta t$, as expected. Based on these
results and analogous calculations for other values of the correlation exponent $\alpha$, we have chosen a time step
of $\Delta t =0.01$ for the majority of our simulations. As a further test, we determine how the
probability density $P(x,t)$ is affected by the time step $\Delta t$. The main panel of Fig.\
\ref{fig:v2vsdt} presents results for $\alpha=1.5$ and $t=20972$. They show that the probability
densities for time steps $\Delta t=0.01$, 0.005, and 0.0025 agree very well,
giving us further confidence in using $\Delta t=0.01$.

Using the above numerical method, we perform simulations for several values
of the correlation exponent $\alpha=1.0, 1.1, 1.2, 1.4, 1.5, 1.6$ and 1.7.
We use up to $N_t=2^{27} \approx 134$ million time steps. All data are averages over a large number of
trajectories ranging between $10^4$  and $5 \times 10^6$.
Unless otherwise noted, our simulations are performed for noise amplitude
$K_\alpha=[\alpha(\alpha-1)]^{-1}$ and a temperature $T=1$. To model the reflecting walls,
we use parameters $F_0=5$ and $\lambda=5$ in eq.\ (\ref{eq:wall_force}).

\section{Finite interval}
\label{sec:finite}

We now turn to the results, starting with simulations of the FLE confined to an
interval $(-L,L)$ by reflecting potentials (\ref{eq:wall_potential}) with the
appropriate signs located at $x_0=\pm L$. In these simulations, the particles start
from rest at the origin, $x=0$, at time $t=0$. We then follow the time evolution
until a steady state is reached.

\begin{figure}
\includegraphics[width=\columnwidth]{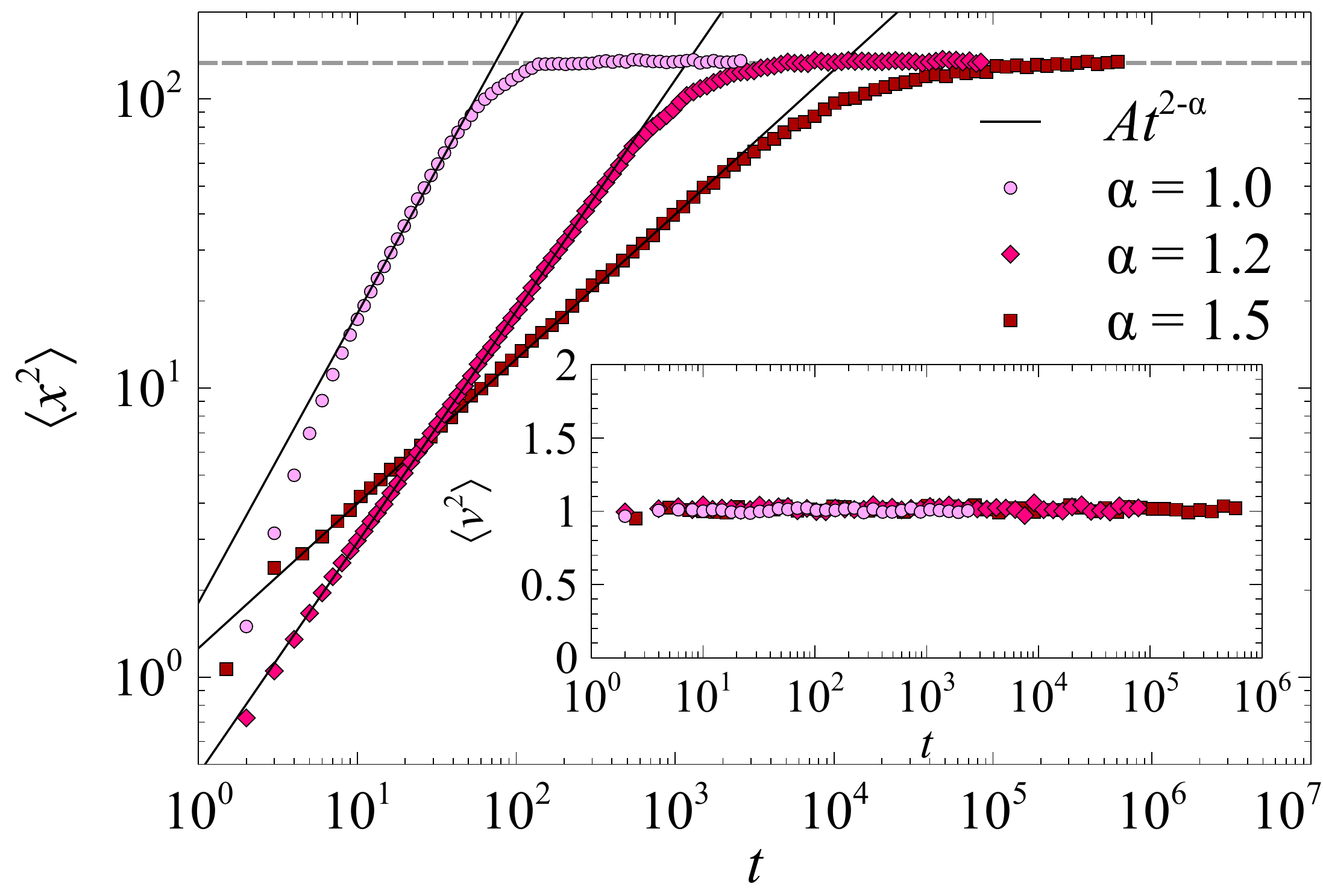}
\caption{FLE confined to the interval $(-L,L)$ with $L=20$. The particles start from rest at the origin, $x=0$, at time $t=0$.
         Main panel: Mean-square displacement $\langle x^2 \rangle$ vs.\ time $t$ for several values of the
         correlation exponent $\alpha$. The data are averages over 10\,000 trajectories. The solid lines are
         fits to the power law (\ref{eq:MSD_free}), $\langle x^2 \rangle \sim t^{2-\alpha}$. The dashed line
         marks the value expected for a uniform distribution of the particles over the interval,
         $L^2/3=400/3$. Inset: Mean-square velocity $\langle v^2 \rangle$ vs.\ time $t$ for the same
         trajectories. }
\label{fig:meansquareinterval}
\end{figure}

Figure \ref{fig:meansquareinterval} presents the time dependence of the mean-square velocity
$\langle v^2 \rangle$ and the mean-square displacement $\langle x^2 \rangle$ for several
values of the correlation exponent $\alpha$ including the uncorrelated case, $\alpha=1$.
For all $\alpha$, the mean-square velocity, shown in the inset, quickly settles on the value 1, as expected
from the fluctuation-dissipation theorem at unit temperature. The mean-square displacement, shown in the main
panel of the figure, features more interesting properties. After transient ballistic behavior at very short times, it grows following
the same power law, $\langle x^2 \rangle \sim t^{2-\alpha}$, as the mean-square displacement
of the free FLE, see eq.\ (\ref{eq:MSD_free}). At long times, it saturates at a constant value that is
independent of the correlation exponent $\alpha$ and coincides with the expectation $\langle x^2 \rangle
= L^2/3$ for a uniform distribution of particles over the interval $(-L,L)$. These results agree with earlier
findings by Jeon et al.\ \cite{JeonMetzler10}.

\begin{figure}
\includegraphics[width=\columnwidth]{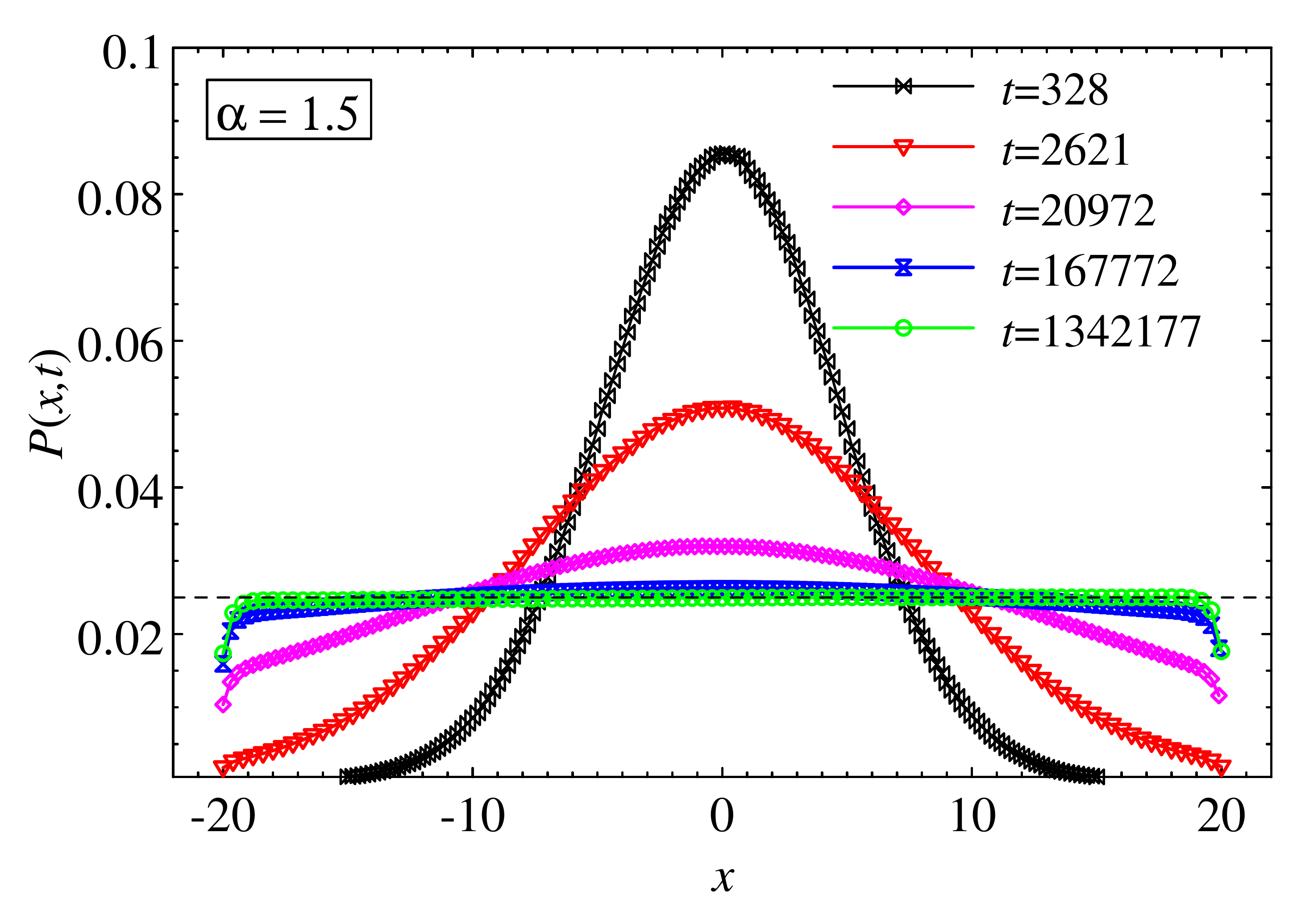}
\caption{Probability density $P(x,t)$ at different times $t$ for the FLE restricted to the interval
         $(-L,L)$ with $L=20$ for $\alpha=1.5$. The data are averages over 10000 trajectories.
         For improved statistics, the data are also averaged over the time interval from $0.8t$ to $t$.
         The dashed line corresponds to a uniform distribution $P(x)=1/(2L)$.}
\label{fig:eq_cond}
\end{figure}

The time evolution of the probability density $P(x,t)$ is shown in Fig.\ \ref{fig:eq_cond} for $\alpha=1.5$.
At short times, $P(x,t)$ is a Gaussian in $x$ centered at the starting point $x=0$ that broadens with time.
Initially, this Gaussian is practically unaffected by the walls.
For longer times, $P(x,t)$ crosses over to a stationary distribution.
In thermal equilibrium, the stationary probability density is expected to follow the Boltzmann distribution,
i.e., $P(x)$ should be proportional to $\exp(-V(x)/T)$. As $V(x)$ vanishes in the interior of the interval,
we expect a uniform distribution. Very close to the wall (at distances of the order of $\lambda^{-1}$) we expect
$P(x)$  to be ``rounded'' due to the wall potential (\ref{eq:wall_potential}).
Specifically, the potential energy of a particle increases upon approaching the wall, suppressing $P(x)$.
As $\lambda$ is a fixed system parameter that does not scale with the interval length $L$,
the rounding becomes negligible in the scaling limit $\lambda L \gg 1$.
The data in Fig.\ \ref{fig:eq_cond} are in perfect agreement with these expectations.
For other values of the correlation exponent
$\alpha$, the probability density $P(x,t)$ behaves in exactly the same way.

We emphasize that the behavior of the FLE on a finite interval observed here is very different from the
behavior of FBM on a finite interval reported in Ref.\ \cite{Guggenbergeretal19}. Whereas the probability density
of the FLE reaches a uniform distribution for long times for all values of the correlation exponent $\alpha$,
the stationary probability density of FBM is not uniform and depends on the value of $\alpha$. (The FBM probability
density is uniform only for uncorrelated noise, $\alpha=1$). This difference is a direct consequence
of the fact that the FLE fulfills the fluctuation-dissipation theorem and can thus reach thermal equilibrium
whereas FBM does not fulfill the fluctuation dissipation theorem (and the fractional Gaussian noise is interpreted
as external noise \cite{Klimontovich_book95}). To test this further, we also perform simulations
of a generalized Langevin equation (\ref{eq:GLE}) with a damping force that does not fulfill the
fluctuation-dissipation relation (\ref{eq:FDT}). This is discussed in Sec.\ \ref{subsec:GLE_without_FDT}.

\section{Semi-infinite interval}
\label{sec:semi-infinite}

\begin{figure}
\includegraphics[width=\columnwidth]{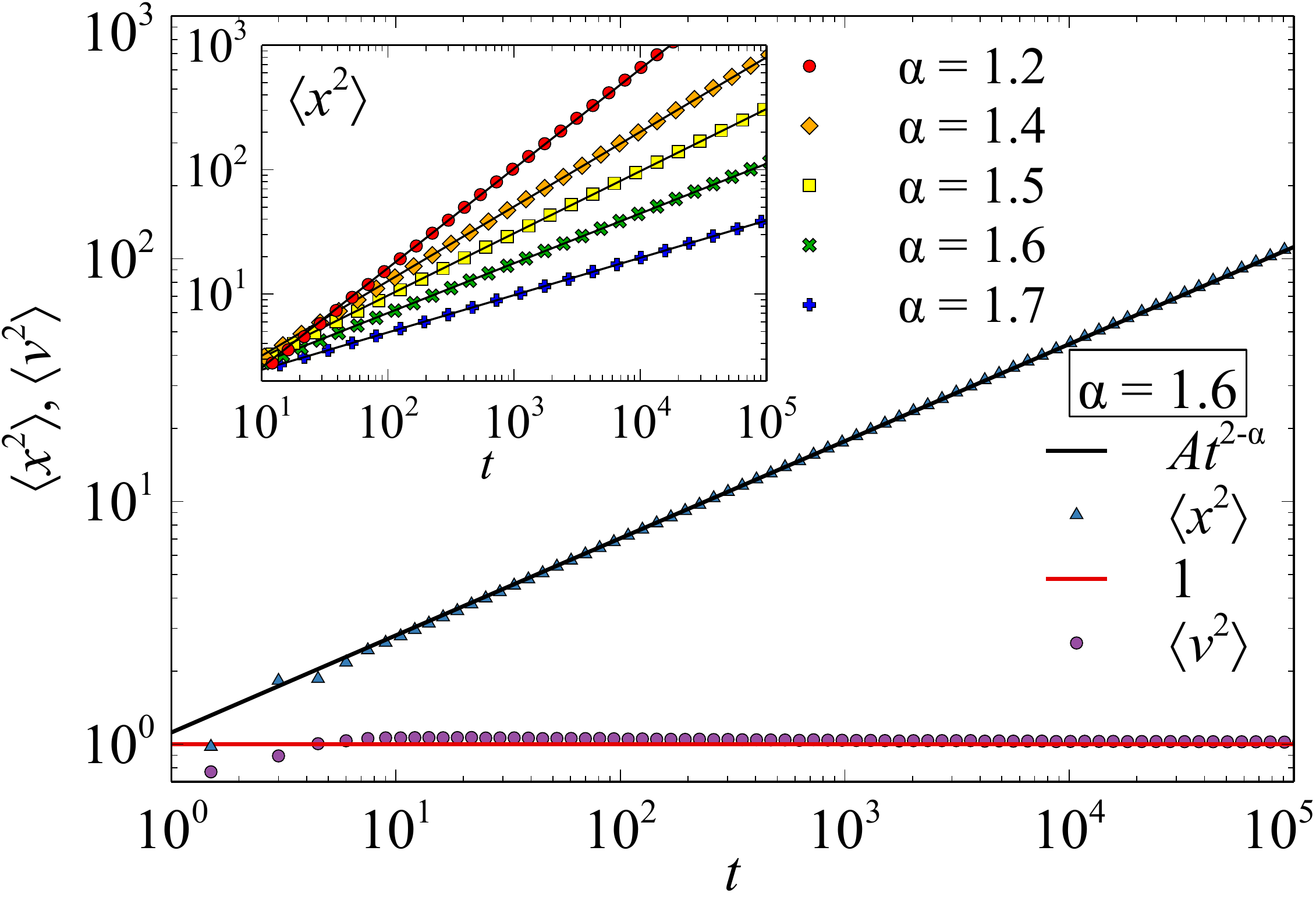}
\caption{FLE on the semi-infinite interval $(0,\infty)$. The particles start from rest at the origin, $x=0$, at time $t=0$.
         Main panel: Mean-square displacement $\langle x^2 \rangle$ and mean-square velocity $\langle v^2 \rangle$ vs.\ time $t$
         for $\alpha=1.6$. The data are averages over $10^6$ trajectories. The (red) horizontal line marks the value
         $\langle v^2 \rangle =1$ expected from the fluctuation dissipation relation. The fit line  for $\langle x^2 \rangle$
         represents a fit to the power law (\ref{eq:MSD_free}), $\langle x^2 \rangle \sim t^{2-\alpha}$.
         Inset: Mean-square displacement $\langle x^2 \rangle$ vs.\ time $t$ for several $\alpha$. The solid lines are
         fits to $\langle x^2 \rangle \sim t^{2-\alpha}$. }
\label{fig:meansquare}
\end{figure}

We now turn to the results for the FLE confined to the semi-infinite interval $(0,\infty)$ by a single reflecting
wall at the origin. Because the motion is unbounded, we expect the mean-square displacement of a particle
that starts at the origin to grow without limit with time.
Figure \ref{fig:meansquare} presents the time evolution of the mean-square displacement $\langle x^2 \rangle$ and
mean-square velocity $\langle v^2 \rangle$ for several values of the correlation exponent $\alpha$.
The figure shows that the mean-square velocity quickly settles on the value $\langle v^2 \rangle = 1$ expected from the
classical equipartition theorem. The mean-square displacement follows the same power law $\langle x^2 \rangle \sim t^{2-\alpha}$
as is observed for the free (unconfined) FLE.

Because the FLE on a semi-infinite interval does not reach a steady state and thus not thermal equilibrium,
the functional form of $P(x,t)$ is not constrained to follow the Boltzmann distribution. In fact, our simulations show
 that $P(x,t)$ has quite unexpected features. For reference, let us first discuss the case of
normal diffusion (which corresponds to $\alpha=1$). The probability density of normal diffusion
with a reflecting wall at the origin can be found by solving the diffusion equation,
$\partial_t P = D \partial_x^2 P$ under the flux-free boundary condition $\partial_x P=0$ at $x=0$.
This yields a Gaussian distribution of the same width as
in the unconfined case but with twice the amplitude (because $P(x,t)$ is restricted to nonnegative values).

\begin{figure}
\includegraphics[width=\columnwidth]{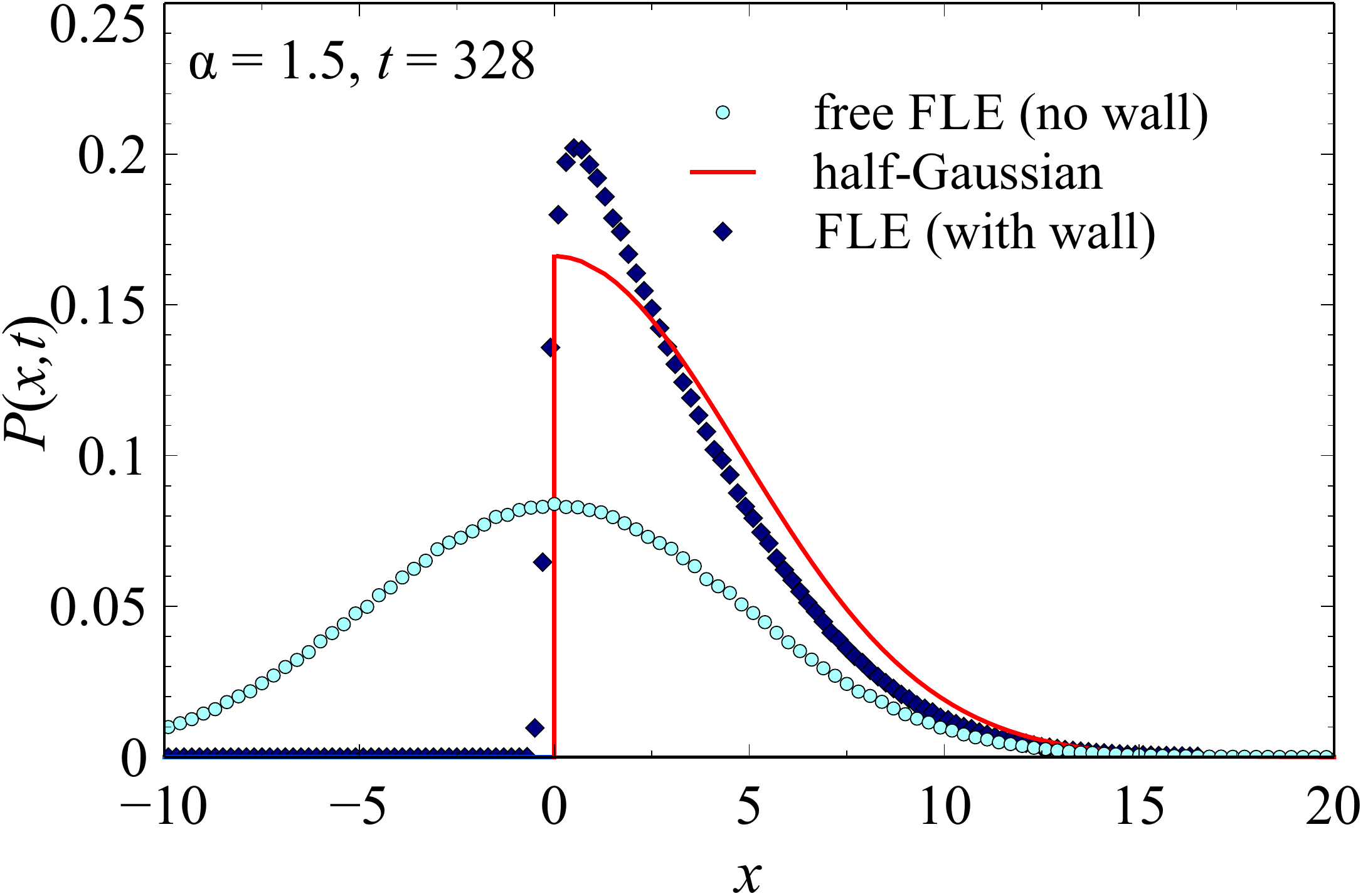}
\caption{Probability densities $P(x,t)$ for the free FLE and the FLE confined to the interval $(0,\infty)$
at time $t=328$ and $\alpha=1.5$. The particles start at $x=0$ at time $t=0$. The data are averages over $10^6$ runs.
The solid line shows the behavior naively expected for the confined case, viz., a half-Gaussian of the same width
but twice the magnitude as the unconfined probability density.}
\label{fig:wall}
\end{figure}

Figure \ref{fig:wall} presents simulation results for the probability densities $P(x,t)$ comparing a free,
unconfined FLE to the FLE on our semi-infinite interval $(0,\infty)$ for correlation exponent $\alpha=1.5$.
(In both cases, the particles start at the origin at $t=0$.)
For the free FLE, the simulations confirm the Gaussian functional form of $P(x,t)$. The probability density on the
semi-infinite interval, however, displays significant deviations from the naively expected half-Gaussian form.
We observe analogous non-Gaussian behavior for all studied $\alpha > 1$.

\begin{figure}
\includegraphics[width=\columnwidth]{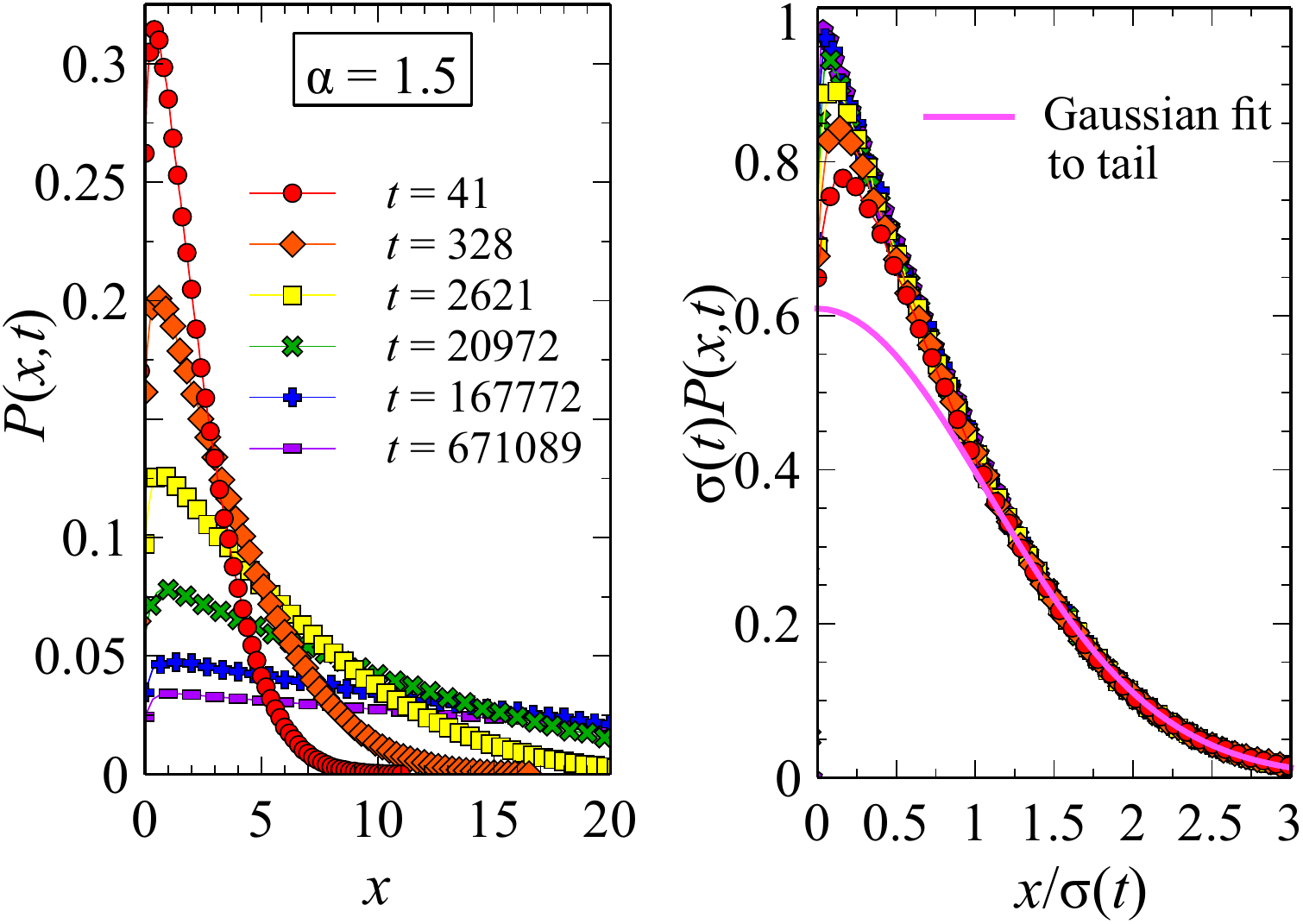}
\caption{Left: Probability densities $P(x,t)$ for the FLE confined to the interval $(0,\infty)$
at different times $t$ for $\alpha=1.5$. The particles start at $x=0$ at time $t=0$. The data are averages over  $10^6$ to
$5\times 10^6$ runs. Right: Scaling plot $\sigma(t) P(x,t)$ vs.\ $x/\sigma(t)$ of the same data. The solid line is a fit of the large-$x$ tail
of the distribution to a Gaussian.}
\label{fig:prob_dist}
\end{figure}

The left panel of Fig.\  \ref{fig:prob_dist} presents the time evolution of the probability density for $\alpha=1.5$.
It shows that the non-Gaussian character persists to the longest times. In fact,
if $x$ is scaled with the root-mean-square displacement $\sigma(t)=\sqrt{\langle x^2(t) \rangle}$ at each time $t$,
the probability densities for all times collapse onto a common master curve, as is
demonstrated in the right panel of Fig.\  \ref{fig:prob_dist}. This implies that the probability
density fulfills the scaling form
\begin{equation}
P(x,t) = \frac 1 {\sigma(t)} Y \left[ \frac x {\sigma(t)} \right] = \frac 1 {b t^{1-\alpha/2}} Y \left[ \frac x {b t^{1-\alpha/2}} \right]
\label{eq:P_scaling}
\end{equation}
where $Y$ is a dimensionless scaling function and $b$ is a constant. The small deviations from a perfect
scaling collapse that can be seen for small $x$ at short times $t$ can be attributed to the ``soft'' wall
potential that rounds $P(x,t)$ over a distance of order $\lambda^{-1}$ close to the wall. This rounding
becomes negligible for long times, i.e., in the scaling regime $\lambda \sigma(t) \gg 1$.

\begin{figure}
\includegraphics[width=\columnwidth]{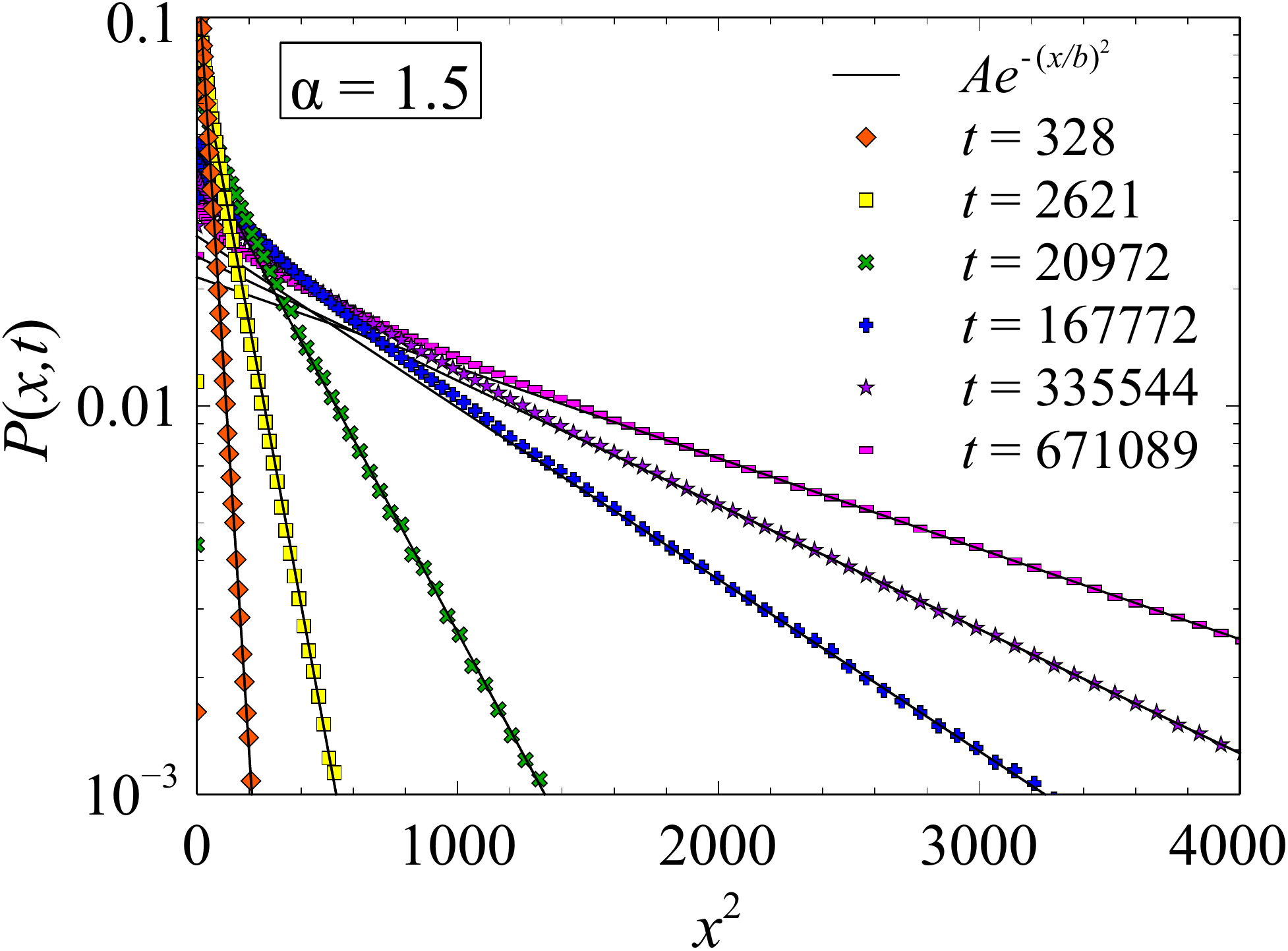}
\caption{Probability density data of Fig.\  \ref{fig:prob_dist}, plotted as $P$ vs.\ $x^2$ such that a Gaussian distribution
gives a straight line. The solid lines are Gaussian fits of the large-$x$ behavior.}
\label{fig:tail}
\end{figure}

The large-$x$ tail of the probability density takes a Gaussian form as can be clearly seen in Fig.\ \ref{fig:tail}
where we replot the data of Fig.\ \ref{fig:prob_dist} as $\log P(x,t)$ vs.\ $x^2$ such that a Gaussian leads to a straight line.
For small $x$, in contrast, the probability density is increased compared to the Gaussian, i.e., particles accumulate at the wall.

\begin{figure}
\includegraphics[width=\columnwidth]{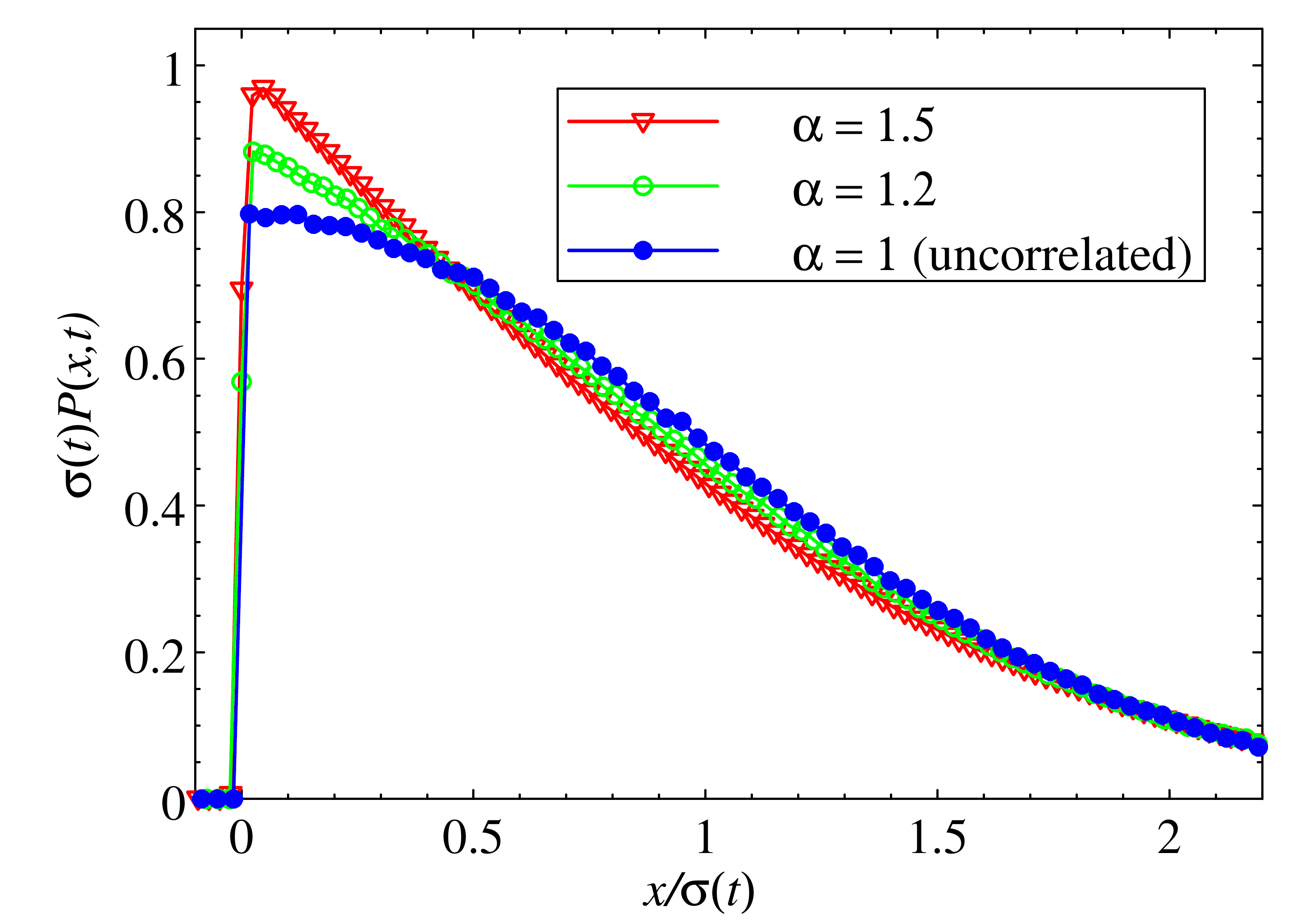}
\caption{Scaled probability density  $\sigma(t) P(x,t)$ vs.\ $x/\sigma$(t) at time
$t=167772$ for several values of the correlation exponent $\alpha$. The data represent averages
of $3 \times 10^6$ trajectories.}
\label{fig:alphas}
\end{figure}

We observe analogous behavior for all values of the correlation exponent $\alpha$ studied in the
range $1 < \alpha <2$. The magnitude of the particle accumulation close to the wall depends on the $\alpha$
value . This is illustrated in Fig.\
\ref{fig:alphas} which shows the scaled probability density at long times for several $\alpha$ values.
For the case of normal diffusion, (uncorrelated random forces, $\alpha=1$) our simulations yield a Gaussian probability density as
expected from the solution of the diffusion equation with a flux-free boundary condition at the origin.
With increasing $\alpha$, i.e., as the correlations become more long-ranged, the deviations of $P(x,t)$ from Gaussian
behavior increase. The exact functional form of $P(x,t)$ at the reflecting wall is hard to determine from the
available numerical data as the rounding due to the soft wall force limits the resolution
close to the wall. However, we see no indications of a divergence for $x/\sigma(t) \to 0$. Instead, the data appear
compatible with a cusp at $x=0$.

It is interesting to compare the non-Gaussian behavior of the probability density of the
FLE on a semi-infinite-interval with the non-Gaussian behavior of reflected FBM studied in Refs.\
\cite{WadaVojta18,WadaWarhoverVojta19}. In both cases, persistent, positively correlated noise
($\alpha >1$) leads to an accumulation of particles at the wall. Moreover, in both cases,
the probability density fulfills the scaling form (\ref{eq:P_scaling}). However, the accumulation
is qualitatively stronger for FBM for which the scaled probability density $\sigma(t) P(x,t)$  diverges
at the wall. This divergence is well described by the power law $P(x,t) \sim x^\kappa$ with $\kappa= 2/\alpha-2$;
it is crucially important in applications that are dominated by rare events (see Ref.\
\cite{WadaSmallVojta18} for an example).
For the FLE, the accumulation of particles at the wall is weaker, and the scaled probability density remains
finite at $x=0$. We emphasize that persistent noise leads to an accumulation at the wall
for both FLE and FBM \emph{despite} the fact that the motion is subdiffusive for the FLE and superdiffusive
for FBM. Subdiffusive FBM which is caused by antipersistent noise, in contrast, leads to a depletion
of particles near the wall. This suggests that the sign of the correlation determines whether or not particles
are accumulated or depleted and not the type of anomalous diffusion (superdiffusion vs.\ subdiffusion).
We note that FBM in a steep confining potential also shows an accumulation of probability
away from the origin (i.e., close to the walls) in the persistent noise case, as compared to
the corresponding Boltzmann distribution \cite{Guggenbergeretal19}.

\section{Extensions}
\label{sec:extensions}
\subsection{Anti-persistent noise}
\label{subsec:anti-persistent}

In the FLE defined by eqs.\ (\ref{eq:GLE}), (\ref{eq:FDT}), and (\ref{eq:FGN}),
the correlation exponent $\alpha$ is restricted to $1 < \alpha <2$, i.e., to the
case of persistent noise. Anti-persistent noise (corresponding to $\alpha$ values
below unity) is impossible because the damping integral diverges for $\alpha < 1$.
However, as the divergence arises from the short-time behavior at $t=t'$, it can be
cut off without changing the physically important long-time behavior of the noise correlations and the
damping kernel. In fact, in the discretized version of the FLE defined in eqs.\
(\ref{eq:FLE_discrete_v}), (\ref{eq:FLE_discrete_x}), (\ref{eq:FGN_discrete}), and
(\ref{eq:FDT_discrete}), the singularity is already cut off as the damping sum
in (\ref{eq:FLE_discrete_v}) remains finite for all correlation exponents in the
physically interesting interval $0 < \alpha < 2$.

In the regime $\alpha<1$, the FLE describes a
peculiar physical situation. As the noise is anti-persistent, the fluctuation-dissipation
theorem implies that the damping kernel ${\mathcal K}_{n-m}$ has negative values
for $n \ne m$. Instead of damping, these terms thus provide anti-damping, i.e., a positive feedback for the velocity.
Consequently, the mean-square displacement in the free unconfined case grows super-diffusively,
$\langle x^2 \rangle \sim t^{2-\alpha}$, for $\alpha < 1$.

To investigate the case of anti-persistent noise, we perform simulations
of the discretized FLE given by (\ref{eq:FLE_discrete_v}), (\ref{eq:FLE_discrete_x})
on the semi-infinite interval $(0,\infty)$ for $\alpha = 0.8$ and $\alpha=0.5$.
We find that the mean-square displacement and the mean-square velocity show the
same qualitative behavior as in the free, unconfined case. This means the mean-square
velocity quickly settles on the value $\langle v^2 \rangle = 1$ prescribed by the
fluctuation-dissipation theorem, and the mean-square displacement grows as
$\langle x^2 \rangle \sim t^{2-\alpha}$.

\begin{figure}
\includegraphics[width=\columnwidth]{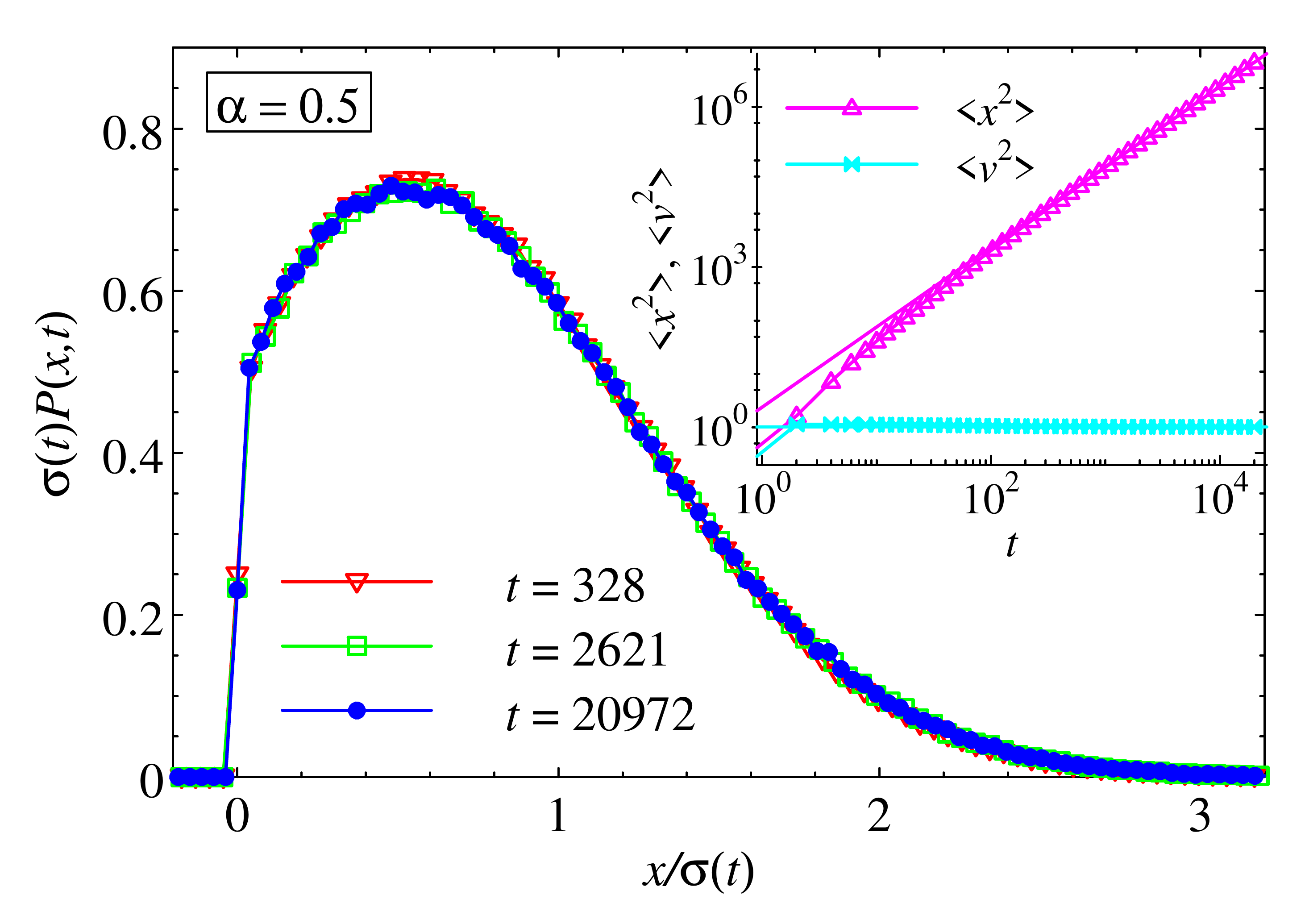}
\caption{Scaled probability density  $\sigma(t) P(x,t)$ vs.\ $x/\sigma(t)$ of particles starting at
$x=0$ at $t=0$ at several times $t$ for $\alpha = 0.5$ and noise amplitude
$K_\alpha=1$. The data represent averages of $10^6$ trajectories.
Inset: Time evolution of the mean-square displacement $\langle x^2 \rangle$ and
mean-square velocity $\langle v^2 \rangle$. The solid lines are fits to
$\langle x^2 \rangle \sim t^{1.5}$ and $\langle v^2 \rangle =\textrm{const}$.
respectively.}
\label{fig:distrib15_scaled}
\end{figure}

The (scaled) probability density of particles starting at the origin at time $t=0$
is shown in Fig.\ \ref{fig:distrib15_scaled} for $\alpha=0.5$.
As in the case of persistent noise, the probability densities taken at different times
collapse when $x$ is scaled with the root-mean-square displacement $\sigma(t)$. This confirms
that the system has reached the asymptotic (scaling) regime at these times. The functional
form of $P(x,t)$ is again non-Gaussian but
in contrast to the case of persistent noise discussed in Sec.\ \ref{sec:semi-infinite},
particles are depleted in the region close to the wall. We observe a similar depletion for
$\alpha=0.8$.

This depletion of particles close to the reflecting wall for the FLE with anti-persistent noise
is analogous to the behavior of FBM with anti-persistent noise \cite{WadaVojta18,WadaWarhoverVojta19}.
However, for FBM, the probability density actually vanishes right at the wall whereas Fig.\
\ref{fig:distrib15_scaled} shows that it remains nonzero for the FLE. As anti-persistent
noise leads to sub-diffusive FBM but super-diffusive motion in the FLE, these results provide
further support for the notion that the sign of the correlation determines whether or not particles
are accumulated or depleted and not the type of anomalous diffusion (superdiffusion vs.\ subdiffusion).

\subsection{Generalized Langevin equation without fluctuation-dissipation theorem}
\label{subsec:GLE_without_FDT}

In Sec.\ \ref{sec:finite}, we found that the probability density
of the FLE on a finite interval reaches a uniform distribution for long times
if the fluctuation-dissipation theorem is fulfilled.
In contrast, the stationary distribution of FBM on a finite interval
is non-uniform \cite{Guggenbergeretal19}. To further test the role that the
fluctuation-dissipation theorem plays in these different situations, we perform simulations
of a generalized Langevin equation (\ref{eq:GLE}) that violates the
fluctuation-dissipation relation (\ref{eq:FDT}).
Specifically, we employ the same long-time correlated
fractional Gaussian noise (\ref{eq:FGN}) as before but combine it with an instantaneous
damping force $-\bar \gamma dx(t)/dt$. (This corresponds to a $\delta$ function
kernel, ${\mathcal K}(t-t')=\delta(t-t')$, in the damping integral in (\ref{eq:GLE}).

\begin{figure}
\includegraphics[width=\columnwidth]{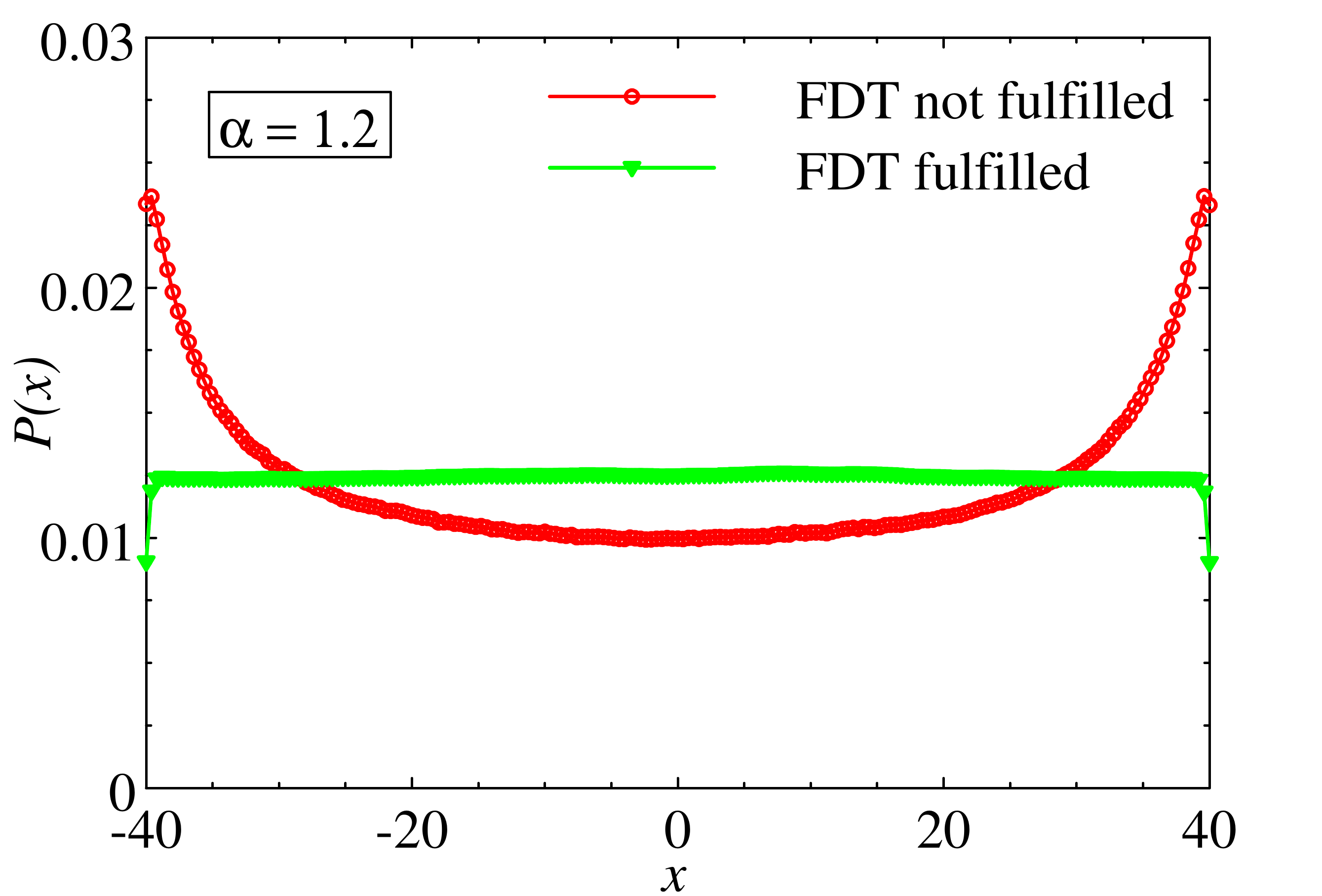}
\caption{Stationary probability density $P(x)$ on the interval $(-L,L)$ with $L=40$ for $\alpha=1.2$.
         Shown are data for the FLE that fulfills the fluctuation dissipation theorem (\ref{eq:FDT}) and a generalized
         Langevin equation with long-time correlated random forces but instantaneous damping that violates it.
         The data are averages over 10000 trajectories.}
\label{fig:comparison}
\end{figure}

Figure\ \ref{fig:comparison} presents a comparison of the stationary probability density of
this generalized Langevin equation with the uniform distribution obtained in Sec.\ \ref{sec:finite}
for the FLE that fulfills the fluctuation-dissipation theorem ($\alpha=1.2$).
The figure demonstrates that the stationary probability density strongly deviates from a uniform distribution if the
fluctuation dissipation theorem is not fulfilled. Instead, it develops singularities close to the walls and
resembles the probability density of reflected FBM found in Ref.\ \cite{Guggenbergeretal19}. We obtained
similar results for $\alpha=1.5$.

In fact, FBM can be understood as the overdamped limit of the generalized Langevin equation (\ref{eq:GLE})
with long-time correlated random forces but instantaneous damping. In the overdamped limit, the acceleration
term can be neglected, and the Langevin equation turns into $\bar \gamma dx(t)/dt = \xi(t)$ which is equivalent to FBM.

\section{Conclusions}
\label{sec:conclusions}

To summarize, motivated by recent observations of a non-Gaussian probability density for reflected
FBM \cite{WadaVojta18,WadaWarhoverVojta19} as well as corresponding deviations from a uniform
stationary probability density for FBM on a finite interval \cite{Guggenbergeretal19},
we studied FLE motion confined by reflecting potential barriers. The main part of our work focuses
on the case when the random and damping forces fulfill the fluctuation-dissipation theorem.

If confined to a finite interval, the FLE therefore reaches thermal equilibrium in the
long-time limit. Correspondingly, our simulations yield a stationary probability density $P(x)$
that is uniform on the interval, independent of the value of the correlation exponent $\alpha$.
This is strikingly different from the behavior of FBM (which contains exactly the same
type of long-range correlated noise) on a finite interval. For FBM, the stationary
probability density is non-uniform and depends on $\alpha$  \cite{Guggenbergeretal19}.

If the FLE is confined to a semi-infinite interval by a single reflecting wall, it will
not reach an equilibrium state because the probability density broadens without limit
(while the mean-square velocity settles on the thermal equilibrium value).
Here, the probability density shows unexpected deviations from the Gaussian form observed for
normal Brownian motion. Specifically, particles accumulate at the reflecting wall for persistent
noise ($1 < \alpha < 2$) while they are depleted close to the wall for anti-persistent
noise ($\alpha <1$). The probability density fulfills the scaling form
$P(x,t) = Y \left[ x /\sigma(t) \right] / \sigma(t) $ where $\sigma(t)$ is the
root-mean-square displacement. This demonstrates that the non-Gaussian behavior is not a
finite-time effect but a feature of the asymptotic long-time behavior.

In other words, the probability density of reflected FBM \cite{WadaVojta18,Guggenbergeretal19}
shows the same type of singular behavior on a finite interval and on a semi-infinite interval,
whereas the FLE shows the usual uniform probability density on a finite interval but
unconventional behavior on a semi-infinite interval.

Let us contrast the behaviors of FBM and FLE motion on a semi-infinite interval. For both
processes, the probability density $P(x,t)$ shows deviations from the Gaussian behavior
observed for normal diffusion. Moreover, for both processes, persistent noise causes
an accumulation of particles at the wall whereas anti-persistent noise leads to a
depletion. This happens even though persistent noise leads to superdiffusive FBM but
subdiffusive motion in the FLE (and vice verse for anti-persistent noise). This suggests
that the behavior of $P(x,t)$ at the wall is controlled by the character of the noise
rather than the type of anomalous diffusion. However, the accumulation and depletion
effects are qualitatively stronger for FBM (where they cause the scaled probability density
$\sigma(t) P(x,t)$ to either vanish or diverge at the wall) than for the FLE (where
$\sigma(t) P(x,t)$ remains finite at the wall).

Our simulations have focused on the unbiased case for which the random force has
zero mean, $\langle \xi(t) \rangle = 0$. However, the expected behavior in the biased
case, $\langle \xi(t) \rangle \ne 0$, is easily discussed. For definiteness, consider
the semi-infinite interval $(0,\infty)$. A bias towards the wall, $\langle \xi(t) \rangle < 0$
corresponds to a potential energy $V(x)$ that increases linearly with the distance from the wall.
In this case, the system is expected to reach thermal equilibrium in the long-time limit.
The stationary probability density is thus given by the equilibrium density
$P(x) \sim \exp[-V(x)/(k_B T)]$ independent of the value of $\alpha$.
If the bias is away from the wall, $\langle \xi(t) \rangle > 0$, the peak of the probability
density will move away from the wall with constant speed whereas its width increases more
slowly. For long times, we will therefore recover the Gaussian probability density of
the unconfined FLE. This means neither of the biased cases is expected to feature an
unusual $\alpha$-dependent probability density in the long-time limit.

To clarify the role that the fluctuation-dissipation theorem plays in establishing the
probability density, we also performed simulations for a generalized Langevin equation
for which the random force $\xi(t)$ is a (long-time correlated) fractional Gaussian noise
but the damping force has no memory and is proportional to the instantaneous velocity.
For this equation, which violates the fluctuation-dissipation theorem, our simulations
show that the stationary probability density on a finite interval is not uniform
but resembles the $\alpha$ dependent probability density observed for FBM.

Recently, Holmes \cite{Holmes19} performed simulations of an overdamped FLE on a finite interval.
In contrast to our results and in violation of the Boltzmann distribution expected
in thermal equilibrium, he reports that the steady-state density is not uniform but
develops depletion zones close to the boundaries. We believe that this discrepancy may
stem from the fact that in Ref.\ \cite{Holmes19} the simple reflection condition
$x_n \to 2 x_0 -x_n, v_n \to -v_n$ is used instead of reflecting potentials to implement the walls.
This effect, especially the role of the simulation time step, is discussed in more detail in Appendix \ref{sec:appendix_A}.

We also note that free FBM and FLE motion are ergodic processes \cite{DengBarkai09,SchwarzlGodecMetzler17}.
FBM and FLE motion were, however, shown to exhibit transient non-ergodic behavior in
a harmonic external potential \cite{JeonMetzler12,Jeonetal13} as well as transient ageing \cite{KursaweSchulzMetzler13}.

Our results show that the probability density functions of FLE motion in a semi-infinite domain, as well as FBM in
confinement and a semi-infinite domain have non-zero slopes at reflecting
boundaries. This is a strong indication for why the traditional method of images \cite{Redner_book01}
fails for processes fueled by correlated fractional Gaussian noise.
Indeed, for FBM, the first passage behavior is only known analytically in the
long time limit in a semi-infinite domain \cite{Molchan99}, in terms of perturbation
theory \cite{Wiese19}, or through conjectures \cite{JeonChechkinMetzler11,SandersAmbjornsson12}.

It will therefore be interesting to compare the reflecting walls considered in this paper with absorbing walls.
Absorbing walls prevent the FLE from reaching a thermal equilibrium state, even on a finite interval.
The probability density is thus not simply given by the appropriate Boltzmann distribution.
Instead, similar to the results of Sec.\ \ref{sec:semi-infinite}, the probability density may
be expected to show nontrivial $\alpha$-dependent behavior. Note that absorbing walls in FBM
are known to produce a power-law singularity of the probability density,
$P(x,t) \sim x^{\kappa}$ similar to reflecting walls but with an exponent $\kappa = 2/\alpha -1 $
\cite{ChatelainKantorKardar08,ZoiaRossoMajumdar09,WieseMahumdarRosso11}
(whereas the exponent takes the value $\kappa = 2/\alpha -2$ at a reflecting wall).

Both the FLE and FBM are used to model a wide variety of anomalous diffusion processes
in complex systems ranging from electronic networks to the motion inside biological cells.
Our results indicate that thermal equilibrium plays a crucial role in the accumulation or depletion
of particles due to the interplay of the long-time memory and the reflecting walls.
In a true equilibrium state, no such accumulation/depletion zones occur. Instead the
probability density follows the Boltzmann distribution, as expected. If, on the other hand,
the system violates the fluctuation-dissipation relation (as is the case for FBM or the
generalized Langevin equation of Sec.\ \ref{subsec:GLE_without_FDT}), the probability density
develops power-law singularities close to the walls and becomes $\alpha$-dependent.
Our simulations also show that correlation-driven accumulation or depletion of particles
can occur even in systems that fulfill the fluctuation dissipation theorem if they are not
in an equilibrium state (as is the case on a semi-infinite interval).
It is interesting to consider this question for biological systems. Some anomalous diffusion
processes in cells are likely well described by thermal equilibrium, others may be dominated
by the active motion of some of the constituents. Working out in detail how this affects
the probability densities remains a task for the future.

\begin{acknowledgments}
This work was supported in part by the NSF under Grant Nos. DMR-1506152 and DMR-1828489 (T.V.)
and by the DFG under Grant No. ME 1535/7-1 (R.M.).  R.M. also acknowledges support
from the Foundation of Polish Science (Fundacja na rzecz Nauki Polskiej) in terms
of an Alexander von Humboldt Polish Honorary Research Scholarship.

\end{acknowledgments}

\appendix
\section{Alternative implementation of reflecting walls}
\label{sec:appendix_A}

\begin{figure}
\includegraphics[width=\columnwidth]{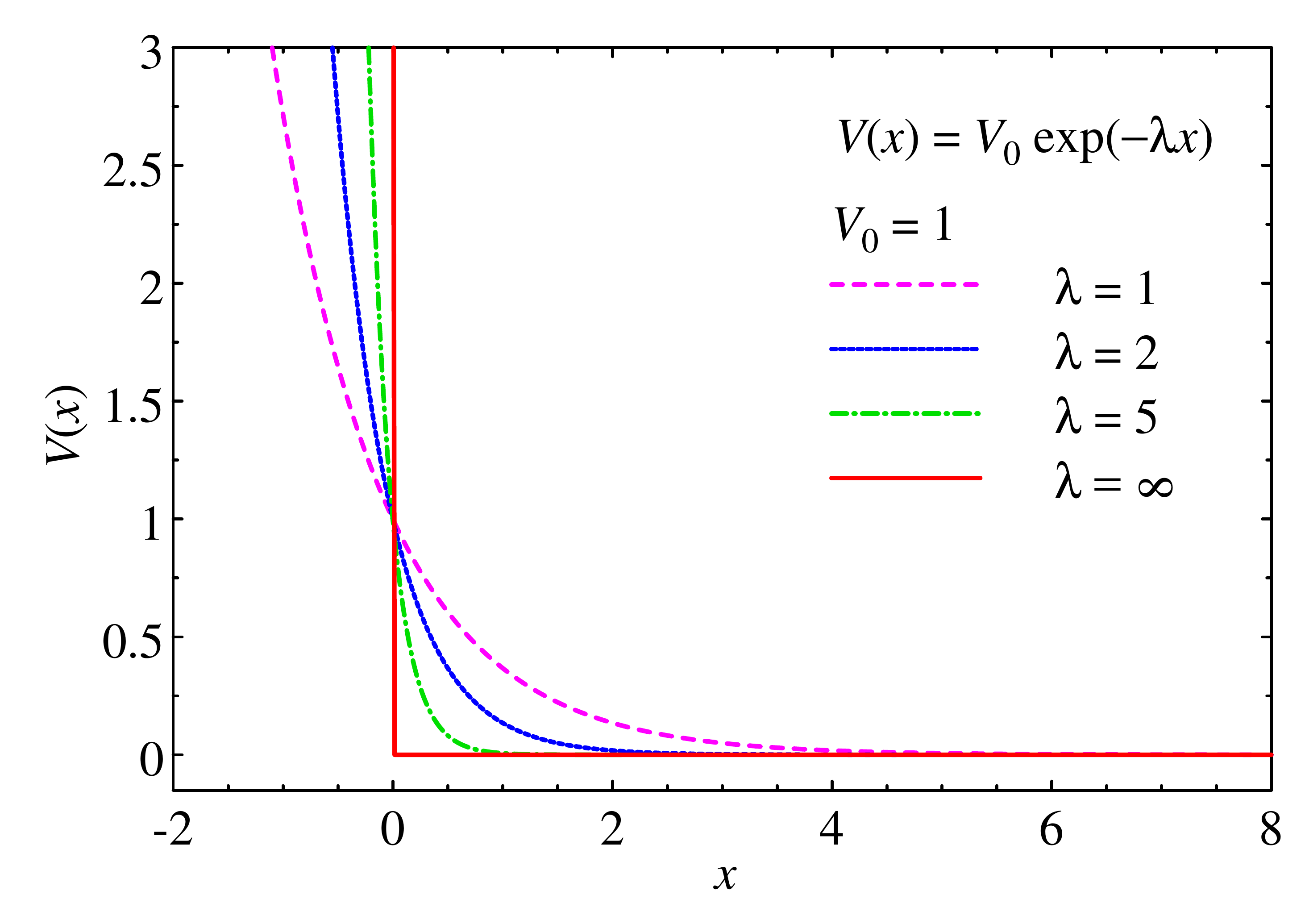}
\caption{Repulsive potential $V(x)$ modeling a reflecting wall at the origin, $x=0$,
for several values of the decay constant $\lambda$.}
\label{fig:potential}
\end{figure}

In the main part of this paper, the reflecting walls have been modeled as ``soft'' repulsive
potentials (\ref{eq:wall_potential}), $V(x)= V_0 \exp[\mp \lambda (x-x_0)]$, as visualized in
Fig.\ \ref{fig:potential}.
Our simulations in Sec.\ \ref{sec:finite} demonstrated that particles governed by
the fractional Langevin equation and confined to a finite interval by such potentials
reach a true thermodynamic equilibrium state. Their kinetic energy fulfills the
equipartition theorem and the spatial probability density follows the Boltzmann
distribution, leading to a uniform stationary $P(x)$.

With increasing $\lambda$ the repulsive potential becomes steeper, reaching an infinitely high
``hard'' wall at position $x_0$ in the limit of $\lambda \to \infty$. Such a wall is expected to
reflect the particle elastically, i.e., the velocity simply changes sign when the particle hits
the wall. In the simulations, it is thus tempting to replace the external force $F(x_n)$ in the
recursion relations (\ref{eq:FLE_discrete_v}) and (\ref{eq:FLE_discrete_x}) by the simple
reflection condition
\begin{equation}
x_n \to 2 x_0 -x_n, \qquad v_n \to -v_n
\label{eq:reflection}
\end{equation}
if the particle finds itself behind a wall located at $x_0$.

Even though this reflection step cannot be interpreted as a normal Euler-like time-step
in the numerical approximation of the fractional Langevin equation (because the change of
velocity is not small), one might still expect it to capture the relevant physics.
In fact, for the normal Langevin equation (\ref{eq:LE}), this implementation of the
reflecting walls leads to the expected behavior.

\begin{figure}
\includegraphics[width=\columnwidth]{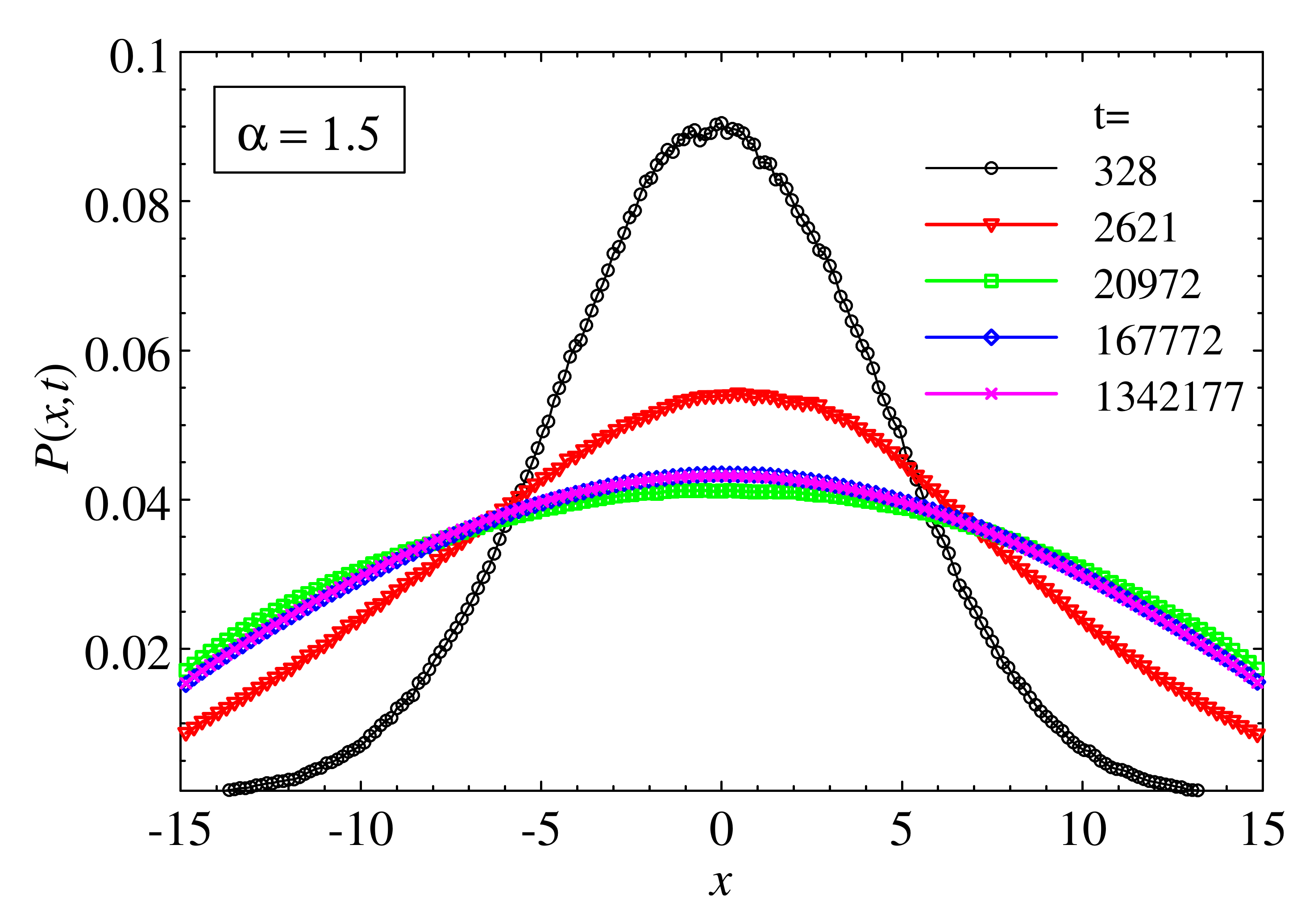}
\caption{Probability density $P(x,t)$ at different times $t$ for the FLE restricted to the interval
         $(-L,L)$ with $L=15$ for $\alpha=1.5$. The data are averages over 4000 trajectories.
        The reflecting walls are implemented via the reflection condition (\ref{eq:reflection}).
        The time step $\Delta t=0.01$, as in the main part of the paper.}
\label{fig:distrib05_interval_hard}
\end{figure}

To test the effects of such hard reflecting boundary conditions on the fractional Langevin equation, we have
analyzed the time evolution of the probability density $P(x,t)$ on a finite interval $(-L,L)$,
combining the recursion relations (\ref{eq:FLE_discrete_v}) and (\ref{eq:FLE_discrete_x})
with the reflection condition (\ref{eq:reflection}) for the two walls at $x=\pm L$.
The results for $L=15$ and a correlation exponent $\alpha=1.5$ are presented in Fig.\ \ref{fig:distrib05_interval_hard}.
The figure shows that the probability density reaches a stationary state for long times, but the stationary
$P(x)$ is not uniform, in contrast to what is expected in thermal equilibrium.
For comparison, the corresponding system with reflecting potentials (\ref{eq:wall_potential}) reaches a uniform
stationary density, as shown in Fig.\ \ref{fig:eq_cond}.

\begin{figure}
\includegraphics[width=\columnwidth]{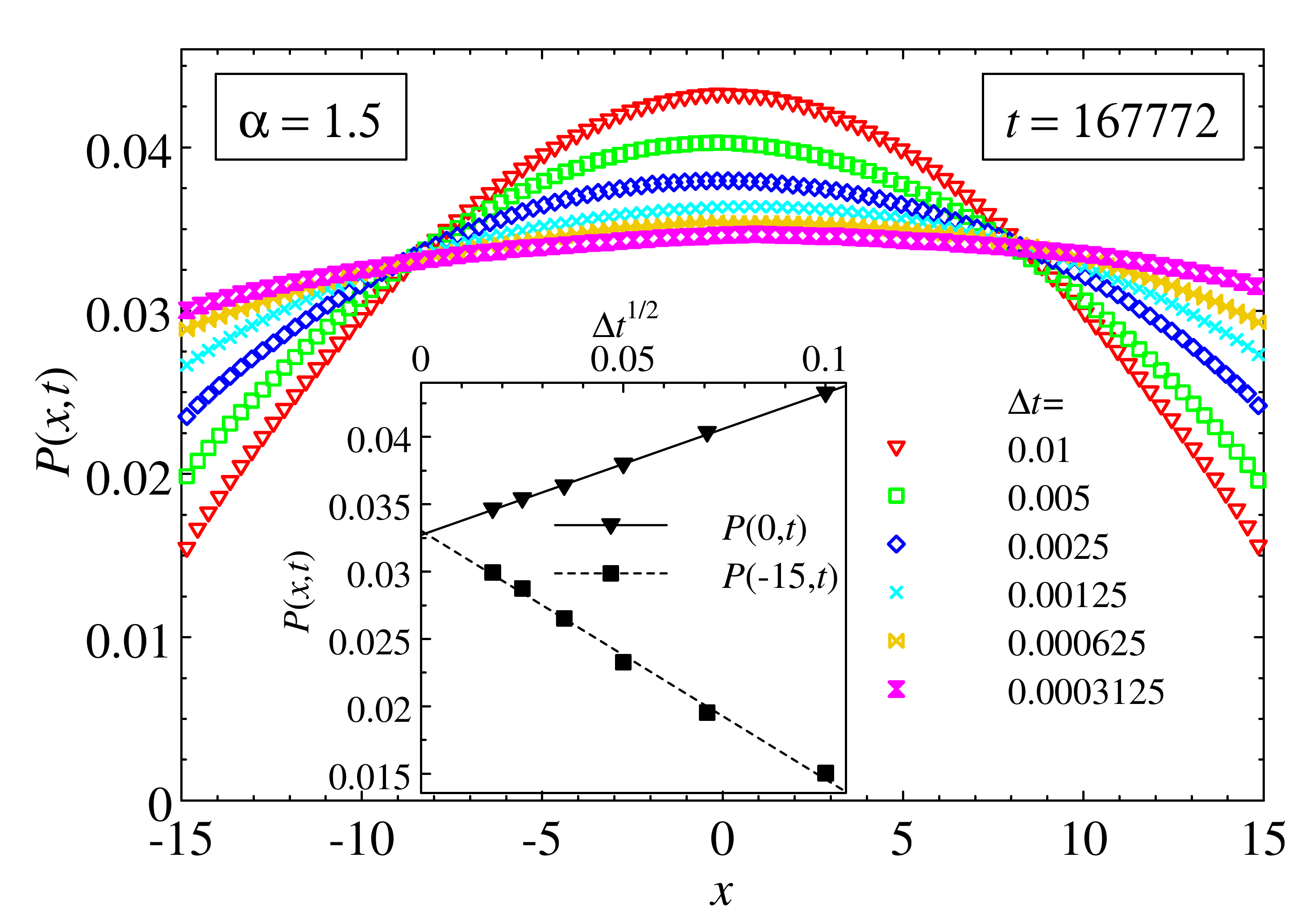}
\caption{Probability density $P(x,t)$ at time $t=167772$ for the FLE on the interval $(-15,15)$
        for several values of the integration time step $\Delta t$. The data are averages over
        4000 to 10000 trajectories. The reflecting walls are implemented via the reflection condition (\ref{eq:reflection}).
        Inset: Extrapolation of $P(0,t)$ and $P(-15,0)$ to time step $\Delta t = 0$.}
\label{fig:hard_delt}
\end{figure}

The data in Fig.\ \ref{fig:distrib05_interval_hard} were obtained using the same
time step $\Delta t =0.01$ that was used successfully in the main part of the paper.
To gain additional insight, we have performed simulations of the FLE with reflection
condition (\ref{eq:reflection}) with several smaller time steps. The resulting probability
densities at a fixed time $t=167772$ are presented in Fig.\ \ref{fig:hard_delt}.
The figure shows that the probability density continues to change with $\Delta t$ down to the smallest studied value of
 $\Delta t = 3.125 \times 10^{-4}$. This needs to be contrasted with the results for the reflecting potential
 shown in Fig.\ \ref{fig:v2vsdt} which are essentially independent of the time step for $\Delta t \lesssim 0.01$.
The deviations in Fig.\ \ref{fig:hard_delt} from a uniform distribution decrease with decreasing $\Delta t$.
The inset of Fig.\ \ref{fig:hard_delt} shows that the probability density $P(0,t)$ in the center of the interval
and $P(-15,t)$ at the wall vary linearly with $(\Delta t)^{1/2}$. We have used this relation to extrapolate $P(0,t)$ and $P(-15,t)$ to zero time step.
Both probability densities extrapolate to the same value which is very close to 0.0333 expected for a uniform distribution.
This suggests that the system actually reaches the equilibrium state in the limit of $\Delta t \to 0$
even if the reflection condition (\ref{eq:reflection}) is used.
However, extremely small $\Delta t$ appear to be necessary to approach that limit. This would significantly
reduce practical applications.

 We have also performed simulations on a semi-infinite interval using the reflection condition
 (\ref{eq:reflection}) and observed similar deviations from the expected behavior. A detailed
 microscopic analysis of this phenomenon will be presented elsewhere
 \footnote{T.\ Vojta, S.\ Skinner, and R.\ Metzler, unpublished.}.

\section{Evaluation of the damping integral}
\label{sec:appendix_B}

A straight-forward implementation of the recursion relations (\ref{eq:FLE_discrete_v})
and (\ref{eq:FLE_discrete_x}) is not very efficient numerically, as the total computational
effort scales quadratically with the number $N_t$ of time steps. This is caused by the damping
term,
\begin{equation}
S_n = \sum_{m=0}^n {\mathcal K}_{n-m} v_m ~,
\label{eq:damping}
\end{equation}
which requires summing over all time steps preceding the current one in each evaluation
of the damping force. The unfavorable
quadratic scaling severely limits the maximum possible simulation times. (Note that
the creation of the correlated random numbers via the Fourier filtering method shows
a much more favorable $N_t \log N_t$ scaling with the number of time steps.)

We have developed an improved method that speeds up the evaluation of the damping sums
by several orders of magnitude. It is based on the observation that, for large
time lag $(n-m)$, the kernel ${\mathcal K}_{n-m}$ is small and slowly varying. This allows
us to approximate it via Taylor expansion. Consider a subsequence of $2j+1$
consecutive terms in the damping sum, centered at term $i$,
\begin{equation}
I_n(i,j)=\sum_{m=i-j}^{i+j} {\mathcal K}_{n-m} v_m~.
\label{eq:partial_sum}
\end{equation}
We now Taylor-expand the kernel about $m=i$ giving
\begin{equation}
{\mathcal K}_{n-m} = {\mathcal K}_{n-i} - (m-i){\mathcal K}_{n-i}^{'} \pm \ldots
\label{eq:Kernel_expansion}
\end{equation}
Inserting this into the partial sum (\ref{eq:partial_sum}) yields
\begin{equation}
I_n(i,j)={\mathcal K}_{n-i} \sum_{m=i-j}^{i+j}  v_m~
         -{\mathcal K}_{n-i}^{'} \sum_{m=i-j}^{i+j}  (m-i) v_m \pm \ldots ~.
\label{eq:partial_sum_expanded}
\end{equation}
The key insight is that the kernel ${\mathcal K}$ and its derivatives can be
precalculated, and the sums in the expression (\ref{eq:partial_sum_expanded})
need to be computed only once and can then be used in every damping sum
at later times. This reduces the numerical effort for computing the partial
sum $I_n(i,j)$ from $2j+1$ multiplications and additions to only one multiplication
and addition if only the leading term in the expansion is kept, or to
two, or perhaps three multiplications and additions if higher order terms
are kept. The accuracy of this approximation can be easily controlled by
varying the number of terms kept in the Taylor expansion and the
width $2j+1$ of the interval used in each partial sum.

In our simulations, we vary the interval width with increasing lag time
$n-m$. Specifically, when calculating the damping sum $S_n$, we calculate
the first $M_0$ terms ($n-m$ from 1 to $M_0$) exactly. We then use intervals
of width $w_1$ until $n-m$ reaches $M_1$, then intervals of width $w_2$
until $n-m$ reaches $M_2$, and so on. Both the widths $w_k$ and the cutoffs
$M_k$ increase in a geometric fashion. For our largest simulations
involving up to $2^{27} \approx 134$ million time steps, we use 4 different interval
widths. (This effectively reduces the scaling of the numerical effort
from $N_t^2$ to $N_t^{1.2}$, albeit with a large prefactor.)
Careful optimization of the interval widths
$w_k$ and cutoff points $M_k$ keeps the relative approximation error for
position and velocity at about $10^{-5}$ while improving the performance by
several orders of magnitude. We have performed systematic benchmarks for a
simulation with $2^{21}$ time steps. On a standard core-i5 PC, the calculation
of a single trajectory takes about 1800 seconds if the damping sums are evaluated
term by term. Our fully optimized version takes only 4 seconds for the same
calculation. For larger numbers of time steps the gains are even bigger.

\bibliographystyle{apsrev4-1}
\bibliography{../../00Bibtex/rareregions}

\begin{thebibliography}{51}%
\makeatletter
\providecommand \@ifxundefined [1]{%
 \@ifx{#1\undefined}
}%
\providecommand \@ifnum [1]{%
 \ifnum #1\expandafter \@firstoftwo
 \else \expandafter \@secondoftwo
 \fi
}%
\providecommand \@ifx [1]{%
 \ifx #1\expandafter \@firstoftwo
 \else \expandafter \@secondoftwo
 \fi
}%
\providecommand \natexlab [1]{#1}%
\providecommand \enquote  [1]{``#1''}%
\providecommand \bibnamefont  [1]{#1}%
\providecommand \bibfnamefont [1]{#1}%
\providecommand \citenamefont [1]{#1}%
\providecommand \href@noop [0]{\@secondoftwo}%
\providecommand \href [0]{\begingroup \@sanitize@url \@href}%
\providecommand \@href[1]{\@@startlink{#1}\@@href}%
\providecommand \@@href[1]{\endgroup#1\@@endlink}%
\providecommand \@sanitize@url [0]{\catcode `\\12\catcode `\$12\catcode
  `\&12\catcode `\#12\catcode `\^12\catcode `\_12\catcode `\%12\relax}%
\providecommand \@@startlink[1]{}%
\providecommand \@@endlink[0]{}%
\providecommand \url  [0]{\begingroup\@sanitize@url \@url }%
\providecommand \@url [1]{\endgroup\@href {#1}{\urlprefix }}%
\providecommand \urlprefix  [0]{URL }%
\providecommand \Eprint [0]{\href }%
\providecommand \doibase [0]{http://dx.doi.org/}%
\providecommand \selectlanguage [0]{\@gobble}%
\providecommand \bibinfo  [0]{\@secondoftwo}%
\providecommand \bibfield  [0]{\@secondoftwo}%
\providecommand \translation [1]{[#1]}%
\providecommand \BibitemOpen [0]{}%
\providecommand \bibitemStop [0]{}%
\providecommand \bibitemNoStop [0]{.\EOS\space}%
\providecommand \EOS [0]{\spacefactor3000\relax}%
\providecommand \BibitemShut  [1]{\csname bibitem#1\endcsname}%
\let\auto@bib@innerbib\@empty
\bibitem [{\citenamefont {Einstein}(1956)}]{Einstein_book56}%
  \BibitemOpen
  \bibfield  {author} {\bibinfo {author} {\bibfnamefont {A.}~\bibnamefont
  {Einstein}},\ }\href@noop {} {\emph {\bibinfo {title} {Investigations on the
  Theory of the Brownian Movement}}}\ (\bibinfo  {publisher} {Dover},\ \bibinfo
  {address} {New York},\ \bibinfo {year} {1956})\BibitemShut {NoStop}%
\bibitem [{\citenamefont {Metzler}\ and\ \citenamefont
  {Klafter}(2000)}]{MetzlerKlafter00}%
  \BibitemOpen
  \bibfield  {author} {\bibinfo {author} {\bibfnamefont {R.}~\bibnamefont
  {Metzler}}\ and\ \bibinfo {author} {\bibfnamefont {J.}~\bibnamefont
  {Klafter}},\ }\href {\doibase https://doi.org/10.1016/S0370-1573(00)00070-3}
  {\bibfield  {journal} {\bibinfo  {journal} {Physics Reports}\ }\textbf
  {\bibinfo {volume} {339}},\ \bibinfo {pages} {1 } (\bibinfo {year}
  {2000})}\BibitemShut {NoStop}%
\bibitem [{\citenamefont {H{\"o}fling}\ and\ \citenamefont
  {Franosch}(2013)}]{HoeflingFranosch13}%
  \BibitemOpen
  \bibfield  {author} {\bibinfo {author} {\bibfnamefont {F.}~\bibnamefont
  {H{\"o}fling}}\ and\ \bibinfo {author} {\bibfnamefont {T.}~\bibnamefont
  {Franosch}},\ }\href {http://stacks.iop.org/0034-4885/76/i=4/a=046602}
  {\bibfield  {journal} {\bibinfo  {journal} {Rep. Progr. Phys.}\ }\textbf
  {\bibinfo {volume} {76}},\ \bibinfo {pages} {046602} (\bibinfo {year}
  {2013})}\BibitemShut {NoStop}%
\bibitem [{\citenamefont {Bressloff}\ and\ \citenamefont
  {Newby}(2013)}]{BressloffNewby13}%
  \BibitemOpen
  \bibfield  {author} {\bibinfo {author} {\bibfnamefont {P.~C.}\ \bibnamefont
  {Bressloff}}\ and\ \bibinfo {author} {\bibfnamefont {J.~M.}\ \bibnamefont
  {Newby}},\ }\href {\doibase 10.1103/RevModPhys.85.135} {\bibfield  {journal}
  {\bibinfo  {journal} {Rev. Mod. Phys.}\ }\textbf {\bibinfo {volume} {85}},\
  \bibinfo {pages} {135} (\bibinfo {year} {2013})}\BibitemShut {NoStop}%
\bibitem [{\citenamefont {Metzler}\ \emph {et~al.}(2014)\citenamefont
  {Metzler}, \citenamefont {Jeon}, \citenamefont {Cherstvy},\ and\
  \citenamefont {Barkai}}]{MJCB14}%
  \BibitemOpen
  \bibfield  {author} {\bibinfo {author} {\bibfnamefont {R.}~\bibnamefont
  {Metzler}}, \bibinfo {author} {\bibfnamefont {J.-H.}\ \bibnamefont {Jeon}},
  \bibinfo {author} {\bibfnamefont {A.~G.}\ \bibnamefont {Cherstvy}}, \ and\
  \bibinfo {author} {\bibfnamefont {E.}~\bibnamefont {Barkai}},\ }\href
  {\doibase 10.1039/C4CP03465A} {\bibfield  {journal} {\bibinfo  {journal}
  {Phys. Chem. Chem. Phys.}\ }\textbf {\bibinfo {volume} {16}},\ \bibinfo
  {pages} {24128} (\bibinfo {year} {2014})}\BibitemShut {NoStop}%
\bibitem [{\citenamefont {Meroz}\ and\ \citenamefont
  {Sokolov}(2015)}]{MerozSokolov15}%
  \BibitemOpen
  \bibfield  {author} {\bibinfo {author} {\bibfnamefont {Y.}~\bibnamefont
  {Meroz}}\ and\ \bibinfo {author} {\bibfnamefont {I.~M.}\ \bibnamefont
  {Sokolov}},\ }\href {\doibase https://doi.org/10.1016/j.physrep.2015.01.002}
  {\bibfield  {journal} {\bibinfo  {journal} {Physics Reports}\ }\textbf
  {\bibinfo {volume} {573}},\ \bibinfo {pages} {1 } (\bibinfo {year}
  {2015})}\BibitemShut {NoStop}%
\bibitem [{\citenamefont {Metzler}\ \emph {et~al.}(2016)\citenamefont
  {Metzler}, \citenamefont {Jeon},\ and\ \citenamefont
  {Cherstvy}}]{MetzlerJeonCherstvy16}%
  \BibitemOpen
  \bibfield  {author} {\bibinfo {author} {\bibfnamefont {R.}~\bibnamefont
  {Metzler}}, \bibinfo {author} {\bibfnamefont {J.-H.}\ \bibnamefont {Jeon}}, \
  and\ \bibinfo {author} {\bibfnamefont {A.}~\bibnamefont {Cherstvy}},\ }\href
  {\doibase https://doi.org/10.1016/j.bbamem.2016.01.022} {\bibfield  {journal}
  {\bibinfo  {journal} {Biochimica et Biophysica Acta}\ }\textbf {\bibinfo
  {volume} {1858}},\ \bibinfo {pages} {2451 } (\bibinfo {year}
  {2016})}\BibitemShut {NoStop}%
\bibitem [{\citenamefont {N{\o}rregaard}\ \emph {et~al.}(2017)\citenamefont
  {N{\o}rregaard}, \citenamefont {Metzler}, \citenamefont {Ritter},
  \citenamefont {Berg-S{\o}rensen},\ and\ \citenamefont
  {Oddershede}}]{Norregaardetal17}%
  \BibitemOpen
  \bibfield  {author} {\bibinfo {author} {\bibfnamefont {K.}~\bibnamefont
  {N{\o}rregaard}}, \bibinfo {author} {\bibfnamefont {R.}~\bibnamefont
  {Metzler}}, \bibinfo {author} {\bibfnamefont {C.~M.}\ \bibnamefont {Ritter}},
  \bibinfo {author} {\bibfnamefont {K.}~\bibnamefont {Berg-S{\o}rensen}}, \
  and\ \bibinfo {author} {\bibfnamefont {L.~B.}\ \bibnamefont {Oddershede}},\
  }\href {\doibase 10.1021/acs.chemrev.6b00638} {\bibfield  {journal} {\bibinfo
   {journal} {Chemical Reviews}\ }\textbf {\bibinfo {volume} {117}},\ \bibinfo
  {pages} {4342} (\bibinfo {year} {2017})}\BibitemShut {NoStop}%
\bibitem [{\citenamefont {Kolmogorov}(1940)}]{Kolmogorov40}%
  \BibitemOpen
  \bibfield  {author} {\bibinfo {author} {\bibfnamefont {A.~N.}\ \bibnamefont
  {Kolmogorov}},\ }\href@noop {} {\bibfield  {journal} {\bibinfo  {journal}
  {Dokl. Acad. Sci. USSR}\ }\textbf {\bibinfo {volume} {26}},\ \bibinfo {pages}
  {115} (\bibinfo {year} {1940})}\BibitemShut {NoStop}%
\bibitem [{\citenamefont {Mandelbrot}\ and\ \citenamefont
  {Ness}(1968)}]{MandelbrotVanNess68}%
  \BibitemOpen
  \bibfield  {author} {\bibinfo {author} {\bibfnamefont {B.~B.}\ \bibnamefont
  {Mandelbrot}}\ and\ \bibinfo {author} {\bibfnamefont {J.~W.~V.}\ \bibnamefont
  {Ness}},\ }\href {\doibase 10.1137/1010093} {\bibfield  {journal} {\bibinfo
  {journal} {SIAM Review}\ }\textbf {\bibinfo {volume} {10}},\ \bibinfo {pages}
  {422} (\bibinfo {year} {1968})}\BibitemShut {NoStop}%
\bibitem [{\citenamefont {Kahane}(1985)}]{Kahane85}%
  \BibitemOpen
  \bibfield  {author} {\bibinfo {author} {\bibfnamefont {J.-P.}\ \bibnamefont
  {Kahane}},\ }\href@noop {} {\emph {\bibinfo {title} {Some Random Series of
  Functions}}}\ (\bibinfo  {publisher} {Cambridge University Press},\ \bibinfo
  {address} {London},\ \bibinfo {year} {1985})\BibitemShut {NoStop}%
\bibitem [{\citenamefont {Yaglom}(1987)}]{Yaglom87}%
  \BibitemOpen
  \bibfield  {author} {\bibinfo {author} {\bibfnamefont {A.~M.}\ \bibnamefont
  {Yaglom}},\ }\href@noop {} {\emph {\bibinfo {title} {Correlation Theory of
  Stationary and Related Random Functions}}}\ (\bibinfo  {publisher}
  {Springer},\ \bibinfo {address} {Heidelberg},\ \bibinfo {year}
  {1987})\BibitemShut {NoStop}%
\bibitem [{\citenamefont {Beran}(1994)}]{Beran94}%
  \BibitemOpen
  \bibfield  {author} {\bibinfo {author} {\bibfnamefont {J.}~\bibnamefont
  {Beran}},\ }\href@noop {} {\emph {\bibinfo {title} {Statistics for
  Long-Memory Processes}}}\ (\bibinfo  {publisher} {Chapman \& Hall},\ \bibinfo
  {address} {New York},\ \bibinfo {year} {1994})\BibitemShut {NoStop}%
\bibitem [{\citenamefont {Biagini}\ \emph {et~al.}(2008)\citenamefont
  {Biagini}, \citenamefont {Hu}, \citenamefont {{\O}ksendal},\ and\
  \citenamefont {Zhang}}]{BHOZ08}%
  \BibitemOpen
  \bibfield  {author} {\bibinfo {author} {\bibfnamefont {F.}~\bibnamefont
  {Biagini}}, \bibinfo {author} {\bibfnamefont {Y.}~\bibnamefont {Hu}},
  \bibinfo {author} {\bibfnamefont {B.}~\bibnamefont {{\O}ksendal}}, \ and\
  \bibinfo {author} {\bibfnamefont {T.}~\bibnamefont {Zhang}},\ }\href@noop {}
  {\emph {\bibinfo {title} {Stochastic Calculus for Fractional Brownian Motion
  and Applications}}}\ (\bibinfo  {publisher} {Springer},\ \bibinfo {address}
  {Berlin},\ \bibinfo {year} {2008})\BibitemShut {NoStop}%
\bibitem [{\citenamefont {Chakravarti}\ and\ \citenamefont
  {Sebastian}(1997)}]{ChakravartiSebastian97}%
  \BibitemOpen
  \bibfield  {author} {\bibinfo {author} {\bibfnamefont {N.}~\bibnamefont
  {Chakravarti}}\ and\ \bibinfo {author} {\bibfnamefont {K.}~\bibnamefont
  {Sebastian}},\ }\href {\doibase
  https://doi.org/10.1016/S0009-2614(97)00075-4} {\bibfield  {journal}
  {\bibinfo  {journal} {Chem. Phys. Lett.}\ }\textbf {\bibinfo {volume}
  {267}},\ \bibinfo {pages} {9 } (\bibinfo {year} {1997})}\BibitemShut
  {NoStop}%
\bibitem [{\citenamefont {Panja}(2010)}]{Panja10}%
  \BibitemOpen
  \bibfield  {author} {\bibinfo {author} {\bibfnamefont {D.}~\bibnamefont
  {Panja}},\ }\href {http://stacks.iop.org/1742-5468/2010/i=02/a=L02001}
  {\bibfield  {journal} {\bibinfo  {journal} {J. Stat. Mech.}\ }\textbf
  {\bibinfo {volume} {2010}},\ \bibinfo {pages} {L02001} (\bibinfo {year}
  {2010})}\BibitemShut {NoStop}%
\bibitem [{\citenamefont {Szymanski}\ and\ \citenamefont
  {Weiss}(2009)}]{SzymanskiWeiss09}%
  \BibitemOpen
  \bibfield  {author} {\bibinfo {author} {\bibfnamefont {J.}~\bibnamefont
  {Szymanski}}\ and\ \bibinfo {author} {\bibfnamefont {M.}~\bibnamefont
  {Weiss}},\ }\href {\doibase 10.1103/PhysRevLett.103.038102} {\bibfield
  {journal} {\bibinfo  {journal} {Phys. Rev. Lett.}\ }\textbf {\bibinfo
  {volume} {103}},\ \bibinfo {pages} {038102} (\bibinfo {year}
  {2009})}\BibitemShut {NoStop}%
\bibitem [{\citenamefont {Jeon}\ \emph
  {et~al.}(2011{\natexlab{a}})\citenamefont {Jeon}, \citenamefont {Tejedor},
  \citenamefont {Burov}, \citenamefont {Barkai}, \citenamefont
  {Selhuber-Unkel}, \citenamefont {Berg-S\o{}rensen}, \citenamefont
  {Oddershede},\ and\ \citenamefont {Metzler}}]{Jeonetal11}%
  \BibitemOpen
  \bibfield  {author} {\bibinfo {author} {\bibfnamefont {J.-H.}\ \bibnamefont
  {Jeon}}, \bibinfo {author} {\bibfnamefont {V.}~\bibnamefont {Tejedor}},
  \bibinfo {author} {\bibfnamefont {S.}~\bibnamefont {Burov}}, \bibinfo
  {author} {\bibfnamefont {E.}~\bibnamefont {Barkai}}, \bibinfo {author}
  {\bibfnamefont {C.}~\bibnamefont {Selhuber-Unkel}}, \bibinfo {author}
  {\bibfnamefont {K.}~\bibnamefont {Berg-S\o{}rensen}}, \bibinfo {author}
  {\bibfnamefont {L.}~\bibnamefont {Oddershede}}, \ and\ \bibinfo {author}
  {\bibfnamefont {R.}~\bibnamefont {Metzler}},\ }\href {\doibase
  10.1103/PhysRevLett.106.048103} {\bibfield  {journal} {\bibinfo  {journal}
  {Phys. Rev. Lett.}\ }\textbf {\bibinfo {volume} {106}},\ \bibinfo {pages}
  {048103} (\bibinfo {year} {2011}{\natexlab{a}})}\BibitemShut {NoStop}%
\bibitem [{\citenamefont {Mikosch}\ \emph {et~al.}(2002)\citenamefont
  {Mikosch}, \citenamefont {Resnick}, \citenamefont {Rootzen},\ and\
  \citenamefont {Stegeman}}]{MRRS02}%
  \BibitemOpen
  \bibfield  {author} {\bibinfo {author} {\bibfnamefont {T.}~\bibnamefont
  {Mikosch}}, \bibinfo {author} {\bibfnamefont {S.}~\bibnamefont {Resnick}},
  \bibinfo {author} {\bibfnamefont {H.}~\bibnamefont {Rootzen}}, \ and\
  \bibinfo {author} {\bibfnamefont {A.}~\bibnamefont {Stegeman}},\ }\href
  {\doibase 10.1214/aoap/1015961155} {\bibfield  {journal} {\bibinfo  {journal}
  {Ann. Appl. Probab.}\ }\textbf {\bibinfo {volume} {12}},\ \bibinfo {pages}
  {23} (\bibinfo {year} {2002})}\BibitemShut {NoStop}%
\bibitem [{\citenamefont {Comte}\ and\ \citenamefont
  {Renault}(1998)}]{ComteRenault98}%
  \BibitemOpen
  \bibfield  {author} {\bibinfo {author} {\bibfnamefont {F.}~\bibnamefont
  {Comte}}\ and\ \bibinfo {author} {\bibfnamefont {E.}~\bibnamefont
  {Renault}},\ }\href@noop {} {\bibfield  {journal} {\bibinfo  {journal} {Math.
  Financ.}\ }\textbf {\bibinfo {volume} {8}},\ \bibinfo {pages} {291} (\bibinfo
  {year} {1998})}\BibitemShut {NoStop}%
\bibitem [{\citenamefont {Rostek}\ and\ \citenamefont
  {Sch{\"o}bel}(2013)}]{RostekSchoebel13}%
  \BibitemOpen
  \bibfield  {author} {\bibinfo {author} {\bibfnamefont {S.}~\bibnamefont
  {Rostek}}\ and\ \bibinfo {author} {\bibfnamefont {R.}~\bibnamefont
  {Sch{\"o}bel}},\ }\href {\doibase
  https://doi.org/10.1016/j.econmod.2012.09.003} {\bibfield  {journal}
  {\bibinfo  {journal} {Econom. Model.}\ }\textbf {\bibinfo {volume} {30}},\
  \bibinfo {pages} {30 } (\bibinfo {year} {2013})}\BibitemShut {NoStop}%
\bibitem [{\citenamefont {Wada}\ and\ \citenamefont
  {Vojta}(2018)}]{WadaVojta18}%
  \BibitemOpen
  \bibfield  {author} {\bibinfo {author} {\bibfnamefont {A.~H.~O.}\
  \bibnamefont {Wada}}\ and\ \bibinfo {author} {\bibfnamefont {T.}~\bibnamefont
  {Vojta}},\ }\href {\doibase 10.1103/PhysRevE.97.020102} {\bibfield  {journal}
  {\bibinfo  {journal} {Phys. Rev. E}\ }\textbf {\bibinfo {volume} {97}},\
  \bibinfo {pages} {020102} (\bibinfo {year} {2018})}\BibitemShut {NoStop}%
\bibitem [{\citenamefont {Wada}\ \emph {et~al.}(2019)\citenamefont {Wada},
  \citenamefont {Warhover},\ and\ \citenamefont {Vojta}}]{WadaWarhoverVojta19}%
  \BibitemOpen
  \bibfield  {author} {\bibinfo {author} {\bibfnamefont {A.~H.~O.}\
  \bibnamefont {Wada}}, \bibinfo {author} {\bibfnamefont {A.}~\bibnamefont
  {Warhover}}, \ and\ \bibinfo {author} {\bibfnamefont {T.}~\bibnamefont
  {Vojta}},\ }\href {\doibase 10.1088/1742-5468/ab02f1} {\bibfield  {journal}
  {\bibinfo  {journal} {J. Stat. Mech.}\ }\textbf {\bibinfo {volume} {2019}},\
  \bibinfo {pages} {033209} (\bibinfo {year} {2019})}\BibitemShut {NoStop}%
\bibitem [{\citenamefont {Guggenberger}\ \emph {et~al.}(2019)\citenamefont
  {Guggenberger}, \citenamefont {Pagnini}, \citenamefont {Vojta},\ and\
  \citenamefont {Metzler}}]{Guggenbergeretal19}%
  \BibitemOpen
  \bibfield  {author} {\bibinfo {author} {\bibfnamefont {T.}~\bibnamefont
  {Guggenberger}}, \bibinfo {author} {\bibfnamefont {G.}~\bibnamefont
  {Pagnini}}, \bibinfo {author} {\bibfnamefont {T.}~\bibnamefont {Vojta}}, \
  and\ \bibinfo {author} {\bibfnamefont {R.}~\bibnamefont {Metzler}},\ }\href
  {\doibase 10.1088/1367-2630/ab075f} {\bibfield  {journal} {\bibinfo
  {journal} {New J. Phys.}\ }\textbf {\bibinfo {volume} {21}},\ \bibinfo
  {pages} {022002} (\bibinfo {year} {2019})}\BibitemShut {NoStop}%
\bibitem [{\citenamefont {Klimontovich}(1995)}]{Klimontovich_book95}%
  \BibitemOpen
  \bibfield  {author} {\bibinfo {author} {\bibfnamefont {Y.~L.}\ \bibnamefont
  {Klimontovich}},\ }\href@noop {} {\emph {\bibinfo {title} {Statistical theory
  of open systems - Volume 1: A unified approach to kinetic description of
  processes in active systems}}}\ (\bibinfo  {publisher} {Kluwer Academic
  Publishers},\ \bibinfo {address} {Dordrecht},\ \bibinfo {year}
  {1995})\BibitemShut {NoStop}%
\bibitem [{\citenamefont {Holmes}(2019)}]{Holmes19}%
  \BibitemOpen
  \bibfield  {author} {\bibinfo {author} {\bibfnamefont {W.~R.}\ \bibnamefont
  {Holmes}},\ }\href {\doibase 10.1016/j.bpj.2019.02.021} {\bibfield  {journal}
  {\bibinfo  {journal} {Biophys. J.}\ }\textbf {\bibinfo {volume} {116}},\
  \bibinfo {pages} {1538} (\bibinfo {year} {2019})}\BibitemShut {NoStop}%
\bibitem [{\citenamefont {Langevin}(1908)}]{Langevin08}%
  \BibitemOpen
  \bibfield  {author} {\bibinfo {author} {\bibfnamefont {P.}~\bibnamefont
  {Langevin}},\ }\href@noop {} {\bibfield  {journal} {\bibinfo  {journal} {C.
  R. Acad. Sci. Paris}\ }\textbf {\bibinfo {volume} {146}},\ \bibinfo {pages}
  {530} (\bibinfo {year} {1908})}\BibitemShut {NoStop}%
\bibitem [{\citenamefont {Zwanzig}(2001)}]{Zwanzig_book01}%
  \BibitemOpen
  \bibfield  {author} {\bibinfo {author} {\bibfnamefont {R.}~\bibnamefont
  {Zwanzig}},\ }\href@noop {} {\emph {\bibinfo {title} {Nonequilibrium
  Statistical Mechanics}}}\ (\bibinfo  {publisher} {Oxford University Press},\
  \bibinfo {address} {Oxford},\ \bibinfo {year} {2001})\BibitemShut {NoStop}%
\bibitem [{\citenamefont {H{\"a}nggi}(1978)}]{Haenggi78}%
  \BibitemOpen
  \bibfield  {author} {\bibinfo {author} {\bibfnamefont {P.}~\bibnamefont
  {H{\"a}nggi}},\ }\href {\doibase 10.1007/BF01351552} {\bibfield  {journal}
  {\bibinfo  {journal} {Z. Phys. B}\ }\textbf {\bibinfo {volume} {31}},\
  \bibinfo {pages} {407} (\bibinfo {year} {1978})}\BibitemShut {NoStop}%
\bibitem [{\citenamefont {Goychuk}(2012)}]{Goychuk12}%
  \BibitemOpen
  \bibfield  {author} {\bibinfo {author} {\bibfnamefont {I.}~\bibnamefont
  {Goychuk}},\ }\enquote {\bibinfo {title} {Viscoelastic subdiffusion:
  Generalized {L}angevin equation approach},}\ in\ \href {\doibase
  10.1002/9781118197714.ch5} {\emph {\bibinfo {booktitle} {Advances in Chemical
  Physics}}}\ (\bibinfo  {publisher} {John Wiley \& Sons, Ltd},\ \bibinfo
  {year} {2012})\ pp.\ \bibinfo {pages} {187--253}\BibitemShut {NoStop}%
\bibitem [{\citenamefont {Kubo}(1966)}]{Kubo66}%
  \BibitemOpen
  \bibfield  {author} {\bibinfo {author} {\bibfnamefont {R.}~\bibnamefont
  {Kubo}},\ }\href {\doibase 10.1088/0034-4885/29/1/306} {\bibfield  {journal}
  {\bibinfo  {journal} {Rep. Progr. Phys.}\ }\textbf {\bibinfo {volume} {29}},\
  \bibinfo {pages} {255} (\bibinfo {year} {1966})}\BibitemShut {NoStop}%
\bibitem [{\citenamefont {Qian}(2003)}]{Qian03}%
  \BibitemOpen
  \bibfield  {author} {\bibinfo {author} {\bibfnamefont {H.}~\bibnamefont
  {Qian}},\ }in\ \href {\doibase 10.1007/3-540-44832-2_2} {\emph {\bibinfo
  {booktitle} {Processes with Long-Range Correlations: Theory and
  Applications}}},\ \bibinfo {editor} {edited by\ \bibinfo {editor}
  {\bibfnamefont {G.}~\bibnamefont {Rangarajan}}\ and\ \bibinfo {editor}
  {\bibfnamefont {M.}~\bibnamefont {Ding}}}\ (\bibinfo  {publisher}
  {Springer},\ \bibinfo {address} {Berlin, Heidelberg},\ \bibinfo {year}
  {2003})\ pp.\ \bibinfo {pages} {22--33}\BibitemShut {NoStop}%
\bibitem [{\citenamefont {Lutz}(2001)}]{Lutz01}%
  \BibitemOpen
  \bibfield  {author} {\bibinfo {author} {\bibfnamefont {E.}~\bibnamefont
  {Lutz}},\ }\href {\doibase 10.1103/PhysRevE.64.051106} {\bibfield  {journal}
  {\bibinfo  {journal} {Phys. Rev. E}\ }\textbf {\bibinfo {volume} {64}},\
  \bibinfo {pages} {051106} (\bibinfo {year} {2001})}\BibitemShut {NoStop}%
\bibitem [{\citenamefont {Makse}\ \emph {et~al.}(1996)\citenamefont {Makse},
  \citenamefont {Havlin}, \citenamefont {Schwartz},\ and\ \citenamefont
  {Stanley}}]{MHSS96}%
  \BibitemOpen
  \bibfield  {author} {\bibinfo {author} {\bibfnamefont {H.~A.}\ \bibnamefont
  {Makse}}, \bibinfo {author} {\bibfnamefont {S.}~\bibnamefont {Havlin}},
  \bibinfo {author} {\bibfnamefont {M.}~\bibnamefont {Schwartz}}, \ and\
  \bibinfo {author} {\bibfnamefont {H.~E.}\ \bibnamefont {Stanley}},\
  }\href@noop {} {\bibfield  {journal} {\bibinfo  {journal} {Phys. Rev. E}\
  }\textbf {\bibinfo {volume} {53}},\ \bibinfo {pages} {5445} (\bibinfo {year}
  {1996})}\BibitemShut {NoStop}%
\bibitem [{Note1()}]{Note1}%
  \BibitemOpen
  \bibinfo {note} {The choice of $\Delta t$ is affected by the magnitude and
  steepness of the wall force (\ref {eq:wall_force}) with larger values of
  $F_0$ and $\lambda $ requiring smaller $\Delta t$.}\BibitemShut {Stop}%
\bibitem [{\citenamefont {Jeon}\ and\ \citenamefont
  {Metzler}(2010)}]{JeonMetzler10}%
  \BibitemOpen
  \bibfield  {author} {\bibinfo {author} {\bibfnamefont {J.-H.}\ \bibnamefont
  {Jeon}}\ and\ \bibinfo {author} {\bibfnamefont {R.}~\bibnamefont {Metzler}},\
  }\href {\doibase 10.1103/PhysRevE.81.021103} {\bibfield  {journal} {\bibinfo
  {journal} {Phys. Rev. E}\ }\textbf {\bibinfo {volume} {81}},\ \bibinfo
  {pages} {021103} (\bibinfo {year} {2010})}\BibitemShut {NoStop}%
\bibitem [{\citenamefont {Wada}\ \emph {et~al.}(2018)\citenamefont {Wada},
  \citenamefont {Small},\ and\ \citenamefont {Vojta}}]{WadaSmallVojta18}%
  \BibitemOpen
  \bibfield  {author} {\bibinfo {author} {\bibfnamefont {A.~H.~O.}\
  \bibnamefont {Wada}}, \bibinfo {author} {\bibfnamefont {M.}~\bibnamefont
  {Small}}, \ and\ \bibinfo {author} {\bibfnamefont {T.}~\bibnamefont
  {Vojta}},\ }\href {\doibase 10.1103/PhysRevE.98.022112} {\bibfield  {journal}
  {\bibinfo  {journal} {Phys. Rev. E}\ }\textbf {\bibinfo {volume} {98}},\
  \bibinfo {pages} {022112} (\bibinfo {year} {2018})}\BibitemShut {NoStop}%
\bibitem [{\citenamefont {Deng}\ and\ \citenamefont
  {Barkai}(2009)}]{DengBarkai09}%
  \BibitemOpen
  \bibfield  {author} {\bibinfo {author} {\bibfnamefont {W.}~\bibnamefont
  {Deng}}\ and\ \bibinfo {author} {\bibfnamefont {E.}~\bibnamefont {Barkai}},\
  }\href {\doibase 10.1103/PhysRevE.79.011112} {\bibfield  {journal} {\bibinfo
  {journal} {Phys. Rev. E}\ }\textbf {\bibinfo {volume} {79}},\ \bibinfo
  {pages} {011112} (\bibinfo {year} {2009})}\BibitemShut {NoStop}%
\bibitem [{\citenamefont {Schwarzl}\ \emph {et~al.}(2017)\citenamefont
  {Schwarzl}, \citenamefont {Godec},\ and\ \citenamefont
  {Metzler}}]{SchwarzlGodecMetzler17}%
  \BibitemOpen
  \bibfield  {author} {\bibinfo {author} {\bibfnamefont {M.}~\bibnamefont
  {Schwarzl}}, \bibinfo {author} {\bibfnamefont {A.}~\bibnamefont {Godec}}, \
  and\ \bibinfo {author} {\bibfnamefont {R.}~\bibnamefont {Metzler}},\
  }\href@noop {} {\bibfield  {journal} {\bibinfo  {journal} {Scientific
  Reports}\ }\textbf {\bibinfo {volume} {7}},\ \bibinfo {pages} {3878}
  (\bibinfo {year} {2017})}\BibitemShut {NoStop}%
\bibitem [{\citenamefont {Jeon}\ and\ \citenamefont
  {Metzler}(2012)}]{JeonMetzler12}%
  \BibitemOpen
  \bibfield  {author} {\bibinfo {author} {\bibfnamefont {J.-H.}\ \bibnamefont
  {Jeon}}\ and\ \bibinfo {author} {\bibfnamefont {R.}~\bibnamefont {Metzler}},\
  }\href {\doibase 10.1103/PhysRevE.85.021147} {\bibfield  {journal} {\bibinfo
  {journal} {Phys. Rev. E}\ }\textbf {\bibinfo {volume} {85}},\ \bibinfo
  {pages} {021147} (\bibinfo {year} {2012})}\BibitemShut {NoStop}%
\bibitem [{\citenamefont {Jeon}\ \emph {et~al.}(2013)\citenamefont {Jeon},
  \citenamefont {Leijnse}, \citenamefont {Oddershede},\ and\ \citenamefont
  {Metzler}}]{Jeonetal13}%
  \BibitemOpen
  \bibfield  {author} {\bibinfo {author} {\bibfnamefont {J.-H.}\ \bibnamefont
  {Jeon}}, \bibinfo {author} {\bibfnamefont {N.}~\bibnamefont {Leijnse}},
  \bibinfo {author} {\bibfnamefont {L.~B.}\ \bibnamefont {Oddershede}}, \ and\
  \bibinfo {author} {\bibfnamefont {R.}~\bibnamefont {Metzler}},\ }\href
  {\doibase 10.1088/1367-2630/15/4/045011} {\bibfield  {journal} {\bibinfo
  {journal} {New J. Phys.}\ }\textbf {\bibinfo {volume} {15}},\ \bibinfo
  {pages} {045011} (\bibinfo {year} {2013})}\BibitemShut {NoStop}%
\bibitem [{\citenamefont {Kursawe}\ \emph {et~al.}(2013)\citenamefont
  {Kursawe}, \citenamefont {Schulz},\ and\ \citenamefont
  {Metzler}}]{KursaweSchulzMetzler13}%
  \BibitemOpen
  \bibfield  {author} {\bibinfo {author} {\bibfnamefont {J.}~\bibnamefont
  {Kursawe}}, \bibinfo {author} {\bibfnamefont {J.}~\bibnamefont {Schulz}}, \
  and\ \bibinfo {author} {\bibfnamefont {R.}~\bibnamefont {Metzler}},\ }\href
  {\doibase 10.1103/PhysRevE.88.062124} {\bibfield  {journal} {\bibinfo
  {journal} {Phys. Rev. E}\ }\textbf {\bibinfo {volume} {88}},\ \bibinfo
  {pages} {062124} (\bibinfo {year} {2013})}\BibitemShut {NoStop}%
\bibitem [{\citenamefont {Redner}(2001)}]{Redner_book01}%
  \BibitemOpen
  \bibfield  {author} {\bibinfo {author} {\bibfnamefont {S.}~\bibnamefont
  {Redner}},\ }\href@noop {} {\emph {\bibinfo {title} {A guide to first-passage
  processes}}}\ (\bibinfo  {publisher} {Cambridge University Press},\ \bibinfo
  {address} {Cambridge},\ \bibinfo {year} {2001})\BibitemShut {NoStop}%
\bibitem [{\citenamefont {Molchan}(1999)}]{Molchan99}%
  \BibitemOpen
  \bibfield  {author} {\bibinfo {author} {\bibfnamefont {G.~M.}\ \bibnamefont
  {Molchan}},\ }\href {\doibase 10.1007/s002200050669} {\bibfield  {journal}
  {\bibinfo  {journal} {Commun. Math. Phys.}\ }\textbf {\bibinfo {volume}
  {205}},\ \bibinfo {pages} {97} (\bibinfo {year} {1999})}\BibitemShut
  {NoStop}%
\bibitem [{\citenamefont {Wiese}(2019)}]{Wiese19}%
  \BibitemOpen
  \bibfield  {author} {\bibinfo {author} {\bibfnamefont {K.~J.}\ \bibnamefont
  {Wiese}},\ }\href {\doibase 10.1103/PhysRevE.99.032106} {\bibfield  {journal}
  {\bibinfo  {journal} {Phys. Rev. E}\ }\textbf {\bibinfo {volume} {99}},\
  \bibinfo {pages} {032106} (\bibinfo {year} {2019})}\BibitemShut {NoStop}%
\bibitem [{\citenamefont {Jeon}\ \emph
  {et~al.}(2011{\natexlab{b}})\citenamefont {Jeon}, \citenamefont {Chechkin},\
  and\ \citenamefont {Metzler}}]{JeonChechkinMetzler11}%
  \BibitemOpen
  \bibfield  {author} {\bibinfo {author} {\bibfnamefont {J.-H.}\ \bibnamefont
  {Jeon}}, \bibinfo {author} {\bibfnamefont {A.~V.}\ \bibnamefont {Chechkin}},
  \ and\ \bibinfo {author} {\bibfnamefont {R.}~\bibnamefont {Metzler}},\ }\href
  {\doibase 10.1209/0295-5075/94/20008} {\bibfield  {journal} {\bibinfo
  {journal} {{EPL} (Europhysics Letters)}\ }\textbf {\bibinfo {volume} {94}},\
  \bibinfo {pages} {20008} (\bibinfo {year} {2011}{\natexlab{b}})}\BibitemShut
  {NoStop}%
\bibitem [{\citenamefont {Sanders}\ and\ \citenamefont
  {Ambj{\"o}rnsson}(2012)}]{SandersAmbjornsson12}%
  \BibitemOpen
  \bibfield  {author} {\bibinfo {author} {\bibfnamefont {L.~P.}\ \bibnamefont
  {Sanders}}\ and\ \bibinfo {author} {\bibfnamefont {T.}~\bibnamefont
  {Ambj{\"o}rnsson}},\ }\href {\doibase 10.1063/1.4707349} {\bibfield
  {journal} {\bibinfo  {journal} {J. Chem. Phys.}\ }\textbf {\bibinfo {volume}
  {136}},\ \bibinfo {pages} {175103} (\bibinfo {year} {2012})}\BibitemShut
  {NoStop}%
\bibitem [{\citenamefont {Chatelain}\ \emph {et~al.}(2008)\citenamefont
  {Chatelain}, \citenamefont {Kantor},\ and\ \citenamefont
  {Kardar}}]{ChatelainKantorKardar08}%
  \BibitemOpen
  \bibfield  {author} {\bibinfo {author} {\bibfnamefont {C.}~\bibnamefont
  {Chatelain}}, \bibinfo {author} {\bibfnamefont {Y.}~\bibnamefont {Kantor}}, \
  and\ \bibinfo {author} {\bibfnamefont {M.}~\bibnamefont {Kardar}},\ }\href
  {\doibase 10.1103/PhysRevE.78.021129} {\bibfield  {journal} {\bibinfo
  {journal} {Phys. Rev. E}\ }\textbf {\bibinfo {volume} {78}},\ \bibinfo
  {pages} {021129} (\bibinfo {year} {2008})}\BibitemShut {NoStop}%
\bibitem [{\citenamefont {Zoia}\ \emph {et~al.}(2009)\citenamefont {Zoia},
  \citenamefont {Rosso},\ and\ \citenamefont {Majumdar}}]{ZoiaRossoMajumdar09}%
  \BibitemOpen
  \bibfield  {author} {\bibinfo {author} {\bibfnamefont {A.}~\bibnamefont
  {Zoia}}, \bibinfo {author} {\bibfnamefont {A.}~\bibnamefont {Rosso}}, \ and\
  \bibinfo {author} {\bibfnamefont {S.~N.}\ \bibnamefont {Majumdar}},\ }\href
  {\doibase 10.1103/PhysRevLett.102.120602} {\bibfield  {journal} {\bibinfo
  {journal} {Phys. Rev. Lett.}\ }\textbf {\bibinfo {volume} {102}},\ \bibinfo
  {pages} {120602} (\bibinfo {year} {2009})}\BibitemShut {NoStop}%
\bibitem [{\citenamefont {Wiese}\ \emph {et~al.}(2011)\citenamefont {Wiese},
  \citenamefont {Majumdar},\ and\ \citenamefont
  {Rosso}}]{WieseMahumdarRosso11}%
  \BibitemOpen
  \bibfield  {author} {\bibinfo {author} {\bibfnamefont {K.~J.}\ \bibnamefont
  {Wiese}}, \bibinfo {author} {\bibfnamefont {S.~N.}\ \bibnamefont {Majumdar}},
  \ and\ \bibinfo {author} {\bibfnamefont {A.}~\bibnamefont {Rosso}},\ }\href
  {\doibase 10.1103/PhysRevE.83.061141} {\bibfield  {journal} {\bibinfo
  {journal} {Phys. Rev. E}\ }\textbf {\bibinfo {volume} {83}},\ \bibinfo
  {pages} {061141} (\bibinfo {year} {2011})}\BibitemShut {NoStop}%
\bibitem [{Note2()}]{Note2}%
  \BibitemOpen
  \bibinfo {note} {T.\ Vojta, S.\ Skinner, and R.\ Metzler,
  unpublished.}\BibitemShut {Stop}%
\end{thebibliography}%
\end{document}